\documentclass[12pt,eadjoint tfnpsf]{article}

\catcode`\@=11
\@addtoreset{equation}{section}

\global\arraycolsep=1pt

\usepackage{amsbsy,amssymb,latexsym,amsfonts, amsmath}
\usepackage{mathrsfs}
\usepackage{graphicx}
\usepackage{youngtab}
\usepackage{marvosym}

\RequirePackage[dvips,usenames]{color}
\definecolor{fireblick}{rgb}{0.698039,0.133333,0.133333}

\newcommand{\beq}{\begin{equation}}
\newcommand{\eeq}{\end{equation}}
\newcommand{\bea}{\begin{eqnarray}}
\newcommand{\eea}{\end{eqnarray}}

\newcommand{\ZZ}{\mathbb{Z}}
\newcommand{\RR}{\mathbb{R}}

\def\Tr{\mathop{\rm Tr}}
\def\E{{\epsilon_{1}}}
\def\EE{{\epsilon_{2}}}
\def\EEE{{\epsilon_3}}
\newcommand\tr{\mathrm{tr}}

\newcommand\adj{\mathrm{adj}}

\renewcommand{\thefootnote}{\fnsymbol{footnote}}

\begin{document}
%
%
\begin{titlepage}

\begin{flushright}
\normalsize
~~~~
SISSA  26/2013/FISI-MATE
\end{flushright}

\vspace{80pt}

\begin{center}
{\LARGE The Stringy Instanton Partition Function}\\
\end{center}

\vspace{25pt}

\begin{center}
{
Giulio Bonelli$^{\heartsuit\spadesuit}$, Antonio Sciarappa$^{\heartsuit}$, Alessandro Tanzini$^{\heartsuit}$ and Petr Vasko$^{\heartsuit}$
}\\
%
\vspace{15pt}
%
$^{\heartsuit}$
International School of Advanced Studies (SISSA) \\via Bonomea 265, 34136 Trieste, Italy 
and INFN, Sezione di Trieste \footnote{email: bonelli,asciara,tanzini,vaskop@sissa.it}\\
\vspace{15pt}
$^{\spadesuit}$
I.C.T.P.\\ Strada Costiera 11, 34014 Trieste, Italy
\end{center}
%
\vspace{20pt}
%
We perform an exact computation of the gauged linear sigma model associated to
a D1-D5 brane system on a resolved $A_1$ singularity.
This is accomplished via
supersymmetric localization on the blown-up two-sphere. 
We show that in the blow-down limit $\mathbb{C}^2/\mathbb{Z}_2$
the partition function reduces to the Nekrasov partition function evaluating the equivariant volume of the instanton
moduli space. 
For finite radius we obtain a tower of world-sheet instanton corrections, that we identify
with the equivariant Gromov-Witten invariants of the ADHM moduli space. We show that these corrections
can be encoded in a deformation of the Seiberg-Witten prepotential.  
From the mathematical viewpoint, the D1-D5 system under study displays a twofold nature:
the D1-branes viewpoint captures the equivariant quantum
cohomology of the ADHM instanton moduli space in the Givental formalism, and the D5-branes viewpoint  
is related to higher rank equivariant Donaldson-Thomas
invariants of $\mathbb{P}^1\times\mathbb{C}^2$.
   


\vfill

\setcounter{footnote}{0}
\renewcommand{\thefootnote}{\arabic{footnote}}

\end{titlepage}

\tableofcontents

\section{Introduction}
\label{sec:intro}

Superstring theory proved to be a really powerful tool to engineer supersymmetric gauge theories and to study
via D-branes 
their properties at a deeper level than the one provided by the perturbative quantum field theoretic definition.
Actually, D-branes theory is richer than its gauge theory low energy limit and provides a larger arena 
to probe quantum space-time geometry as seen by superstrings.

A particularly important step in the study of non perturbative phenomena in four dimensional supersymmetric gauge theories 
with eight supercharges was taken by Nekrasov in \cite{Deng}
paving the way to a microscopic derivation of the celebrated Seiberg-Witten (SW) solution \cite{SW}.
The Nekrasov partition function indeed provides an extension of the SW prepotential including an infinite
tower of gravitational corrections coupled to the parameters of the so called $\Omega$-background.
The ability to resum the multi-instanton series crucially depends on the use of equivariant localization technique,
which lastly became a commonly used technique for the exact evaluation of supersymmetric path integrals.
In most cases this technique allows to reduce the path integration over the infinite dimensional space of field configurations
to a localized sum over the points in the moduli space of BPS configurations which are fixed under the 
maximal torus of the global symmetries of the theory.
In the case of ${\cal N}=2$ theories in four dimensions the supersymmetric partition function actually computes the 
equivariant volume of the instanton moduli space.
From a mathematical viewpoint the Nekrasov partition function encodes the data of the classical equivariant 
cohomology of the ADHM instanton moduli space and computes, in presence of observables, equivariant Donaldson polynomials 
\cite{Naka}.

A D-brane engineering of the pure $SU(N)$ gauge theory is provided by a system of $N$ D3-branes at the 
singular point of the orbifold geometry ${\mathbb C}^2/\ZZ_2$. 
The non-perturbative contributions to this theory are then encoded by  
D(-1)-branes which provide the corresponding instanton contributions \cite{DW,Billo}.
The Nekrasov partition function can indeed be computed from the D(-1)-branes point of view as a supersymmetric $D=0$
path integral whose fields realize the open string sectors of the D(-1)-D3 system \cite{Deng,BFMT}.
A particularly relevant point to us is that the open string sectors correspond to the ADHM data 
and 
the super-potential of the system imposes the ADHM constraints on the vacua.

A richer description of the construction above, which avoids the introduction of fractional D-brane charges, 
is obtained by resolving the orbifold $A_1$ singularity to a smooth ALE space obtained by blowing up the singular 
point to a two-sphere \cite{DDG}.
The resolution generates a local K3 smooth geometry, namely the Eguchi-Hanson space, 
given by the total space of the cotangent bundle to the 2-sphere.
The ${\cal N}=2$ $D=4$ gauge theory is then obtained by considering the system of D1-D5 branes wrapping the blown-up 
2-sphere in the zero radius limit.

The aim of this paper is to study the D1-D5 system on the Eguchi-Hanson space at finite radius 
by defining and computing exactly its partition function and to analyse some
mathematical properties of the latter. 
From the D1-branes perspective,
the theory describing the D1-D5 brane system on the resolved space 
is a gauged linear sigma model (GLSM) on the blown-up two-sphere describing the corresponding open string sectors 
with a superpotential interaction which imposes the ADHM constraints.
The $S^2$ partition function of supersymmetric GLSMs can be exactly computed by equivariant 
localization on the two-sphere along the general analysis proposed in \cite{BC,gomisetal}. We will specify their analysis 
to our case to compute the partition function of the D1-D5 system. 
The infrared dynamics of the GLSM describes a non-linear $(2,2)$ sigma model with target space the ADHM moduli 
space itself.
Therefore, the D1-D5 system probes the ADHM geometry from a stringy point of view.
The supersymmetric sigma model contains stringy instanton corrections corresponding to the topological sectors with non trivial
magnetic flux on the 
two-sphere\footnote{These are effective stringy instantons in the ADHM moduli space which compute the KK corrections due to
the finite size of the blown-up ${\mathbb P}^1$. For the sake of clarity, gravity is decoupled from the D-branes and $\alpha'$ 
is scaled away as usual.}. 
The trivial sector, i.e. the sector of constant maps, 
is the only one surviving the zero radius (i.e. point particle) limit; we will show that it reproduces the Nekrasov partition function.

The supersymmetric partition function we define extends the Nekrasov
partition function by including stringy instanton corrections to the equivariant volume 
of the ADHM moduli space.
This follows as a natural extension of the interpretation of the $S^2$ partition function of the GLSM
in terms of the  
quantum Kahler potential of the NLSM geometry to which it flows to in the infrared.
From the mathematical view point the stringy instantons are therefore deforming the classical cohomology of the ADHM
moduli space to a quantum one. 
As we will briefly discuss in this paper, the supersymmetric localization results have a direct link with
Givental's formalism for equivariant quantum cohomology. 
Actually, this is a much more general subject encompassing both compact and non-compact K\"ahler manifolds; 
we elaborate on this topic
in a separate publication \cite{BSTV}. 
The central object of Givental formalism is given by the so-called ${\cal J}$-function which encodes  
the Gromov-Witten invariants and gravitational descendants of the target space.
In the following we will discuss how our results provide a conjectural expression for Givental's
${\cal J}$-function of the ADHM moduli space, and provide explicit checks of this conjecture
for abelian instantons, whose moduli space ${\cal M}_{k,1}$ is described by the Hilbert scheme of points ${\rm Hilb}^k(\mathbb{C}^2)$,
and for one instanton in $U(N)$ gauge theory, whose moduli space ${\cal M}_{1,N}$ reduces to the cotangent bundle
of the $N-1$ dimensional complex projective space $T^*\mathbb{P}^{N-1}$.

  In Section \ref{sec:glsm} we discuss the ADHM gauged linear sigma model from the D1-D5 system perspective
and the calculation of the partition function via supersymmetric localization on the sphere. In particular we discuss how 
this reproduces the Nekrasov instanton partition function in the point particle limit.
    In Section \ref{sec:GW} we study the relation between the spherical partition function and the 
quantum Kahler potential on the ADHM moduli space. We compare its structure with the Givental formalism, identify 
the vortex partition functions as the Givental's function and discuss how to compute out of it the quantum cohomology 
of the ADHM moduli space. We will check our results to reproduce already known results
in some cases, namely the case ${\cal M}_{k,1}$ of $k$ D1s and a single D5-brane and the case of 
a single D1 and $N$ D5-branes ${\cal M}_{1,N}$.
In Section \ref{sec:DT} we explore the system from the D5-brane perspective and propose a relation with higher rank equivariant Donaldson-Thomas
theory on $\mathbb{P}^1\times\mathbb{C}^2$. We show that the free-energy of the D5-brane theory is a deformation
of the Seiberg-Witten prepotential in the $\Omega$-background containing the whole tower of effective world-sheet instanton
corrections. 
 Finally in Section \ref{sec:conclusion} we present our conclusions and discussions on further directions, 
  and collect some useful identities in the Appendices.

\section{ADHM gauged linear sigma model from the D1-D5 system}
\label{sec:glsm}
 
In this section we describe the dynamics of a system of $k$ D1 and $N$ D5-branes wrapping the blown-up sphere of a resolved $A_1$
singularity. Specifically, we consider the type IIB background $\RR^{1,3}\times T^*\mathbb{P}^1\times\RR^2$ with
the D1-branes wrapping the $\mathbb{P}^1$ and space-time filling D5-branes wrapped on $\mathbb{P}^1$.
We focus on the D1-branes, whose dynamics is described by a two-dimensional ${\cal N}=(2,2)$ 
gauged linear sigma model flowing in the infrared to a non-linear sigma model with target space the ADHM moduli space of instantons $\mathcal{M}_{k,N}$.
The field content is reported in the table below.

\begin{table}[h!]
\begin{center}
\begin{tabular}{c|c|c|c|c|c}
{} & $\chi$ & $B_{1}$ & $B_{2}$ & $I$ & $J$ \\ \hline
D-brane sector & D1/D1 & D1/D1 & D1/D1 & D1/D5 & D5/D1 \\ \hline
gauge $U(k)$ & $Adj$ & $Adj$ & $Adj$ & $\mathbf{k}$ & $\mathbf{\bar{k}}$ \\ \hline
flavor $U(N)\times U(1)^{2}$ & $\mathbf{1}_{(-1,-1)}$ & $\mathbf{1}_{(1,0)}$ & $\mathbf{1}_{(0,1)}$ & $\mathbf{\bar{N}}_{(0,0)}$ & $\mathbf{N}_{(1,1)}$ \\ \hline
twisted masses & $\epsilon$ & $-\epsilon_{1}$ & $-\epsilon_{2}$ & $-a_{i}$ & $a_{j}-\epsilon$ \\ \hline
$R$-charge & $2-2q$ & $q$ & $q$ & $q+p$ & $q-p$ \\ \hline
\end{tabular} 
\caption{ADHM gauged linear sigma model}
\end{center}
\end{table} 

\noindent The superpotential of our model is $W=\textrm{Tr}_{k}\left\{\chi\left([B_{1},B_{2}]+IJ\right)\right\}$. 
It implements as a constraint the fact that an infinitesimal open string plaquette in the D1-D1 sector can be undone as a couple
of open strings stretching from the D1 to a D5 and back.
We also consider twisted masses corresponding to 
the maximal torus in the global symmetry group
$U(1)^{N+2}$ acting on ${\cal M}_{k,N}$ which we denote as $(a_{j},-\epsilon_{1},-\epsilon_{2})$. 
The $R$-charges are assigned as the most general ones which ensures $R(W)=2$
and full Lorentz symmetry at zero twisted masses. These provide an imaginary part
to the twisted masses via the
redefinition
\begin{equation}
a_i - i\frac{p+q}{2} \,\longrightarrow\, a_i \;\;\;,\;\;\; \epsilon_{1,2} - i \frac{q}{2} \,\longrightarrow\, \epsilon_{1,2}
\end{equation} 

The computation of the partition function of the gauged linear sigma model on the two-sphere can be performed via equivariant localization
\cite{BC,gomisetal}. Here we follow the notation of \cite{BC}.
The path integral localization is performed with respect to the supercharge $\mathcal{Q}=Q+Q^{\dagger}$, where $Q=\epsilon^{\alpha}Q_{\alpha}$ and $Q^{\dagger}=-(\epsilon^{\dagger}C)^{\alpha}Q^{\dagger}_{\alpha}$ with $C$ being the charge conjugation matrix. $\epsilon$ is a particular solution to the Killing spinor equation chosen as $\epsilon=e^{i\frac{\phi}{2}}(\cos\frac{\theta}{2},i\sin\frac{\theta}{2})$. The supercharges $Q, Q^{\dagger}$ form a $\mathfrak{su}(1|1)$ subalgebra of the full superalgebra, up to a gauge transformation $G$,
\begin{equation}
\left\{Q,Q^{\dagger}\right\}=M+\frac{R}{2}+iG,\qquad Q^{2}=\left(Q^{\dagger}\right)^{2}=0,
\end{equation}
where $M$ is the generator of isometries of the sphere infinitesimally represented by the Killing vector $v=\left(\epsilon^{\dagger}\gamma^{a}\epsilon\right)e_{a}=\frac{1}{r}\frac{\partial}{\partial\phi}$ and $R$ is the generator of the $U(1)_{R}$ symmetry. The Killing vector field generates $SO(2)$ rotations around the axis fixed by the North and South pole. 
Finally, the localizing supercharge $\mathcal{Q}$ satisfies
\begin{equation}
\mathcal{Q}^{2}=M+\frac{R}{2}+iG.
\end{equation}
The fact that $M$ generates a $U(1)$ isometry with the North and South poles as fixed points will play a r\^ole in Section~\ref{sec:GW}.
For convenience we briefly summarize the field content and the action of the ${\cal N}=(2,2)$ GLSM on $S^{2}$.
By dimensional reduction of ${\cal N}=1$ multiplets in four dimensions we get
\begin{equation}
\begin{split}
\text{vector multiplet:}&\quad\left(A_{\mu},\sigma,\eta,\lambda,\bar{\lambda},D\right)\\
\text{chiral multiplet:}&\quad\left(\phi,\bar{\phi},\psi,\bar{\psi},F,\bar{F}\right).
\end{split}
\end{equation}
The action is 
\begin{equation}
S=
\int\{d^{2}x\}\left(\mathcal{L}_{YM}+\mathcal{L}_{FI+top}+\mathcal{L}_{matter}+\mathcal{L}_{W}\right),
\end{equation}
The expressions for the Lagrange densities are
\begin{equation}
\begin{split}
\mathcal{L}_{YM} = \frac{1}{g^{2}}\Tr \bigg\{ \frac12 \Big( F_{12} - \frac \eta r \Big)^2 + \frac12 \Big( D + \frac\sigma r \Big)^2 &+ \frac12 D_\mu\sigma D^\mu \sigma + \frac12 D_\mu\eta D^\mu \eta - \frac12 [\sigma,\eta]^2 \\
&+ \frac i2 \bar\lambda \gamma^\mu D_\mu \lambda + \frac i2 \bar\lambda [\sigma,\lambda] + \frac12 \bar\lambda \gamma_3 [\eta,\lambda] \bigg\}
\end{split}
\end{equation}
\begin{equation}
\begin{split}
\mathcal{L}_{matter} = &D_\mu \bar\phi D^\mu \phi + \bar\phi \sigma^2 \phi + \bar\phi \eta^2 \phi + i \bar\phi D \phi + \bar F F + \frac{iq}r \bar\phi \sigma \phi + \frac{q(2-q)}{4r^2} \bar\phi \phi \\
&- i \bar\psi \gamma^\mu D_\mu \psi + i \bar\psi \sigma \psi - \bar\psi \gamma_3 \eta \psi + i \bar\psi \lambda \phi - i \bar\phi \bar\lambda \psi - \frac q{2r} \bar\psi \psi
\end{split}
\end{equation}
\begin{equation}
\mathcal{L}_{FI+top} = - i \xi D + i \frac\theta{2\pi} F_{12}
\end{equation}
\begin{equation}
\mathcal{L}_W = \sum_j \frac{\partial W}{\partial\phi_j} F_j - \sum_{j,k} \frac12 \frac{\partial^2 W}{\partial\phi_j \partial \phi_k} \psi_j \psi_k
\end{equation}
with $r$ the radius of the sphere and $q$ the R-charge of the chiral multiplet.
To localize on field configurations corresponding to the Coulomb branch the following $\mathcal{Q}$ exact deformation of the action was chosen
\begin{equation}
\delta S=\int\{d^{2}x\}\left(\mathcal{L}_{YM}+\mathcal{L}_{\psi}\right),
\end{equation}
where
\begin{equation}
\mathcal{L}_{YM} = \Tr \mathcal{Q}\, \frac{(\overline{\mathcal{Q}\lambda}) \lambda + \lambda^\dag (\overline{\mathcal{Q}\lambda^\dag})}4  \;,\qquad\qquad \mathcal{L}_\psi = \mathcal{Q}\, \frac{(\overline{\mathcal{Q}\psi}) \psi + \psi^\dag (\overline{\mathcal{Q}\psi^\dag}) }2 \;.
\end{equation}
This procedure reduces the path integral to an ordinary integral over the constant modes of the scalar field $\sigma$ and a sum over the
non trivial fluxes of the gauge field on the two-sphere.
The master formula for the partition function on $S^{2}$ in terms of a contour integral was obtained from this setting in \cite{BC} and from a similar one in \cite{gomisetal}. We apply it to our specific model.

Our computations are valid for $q>p>0\,,\, q<1$, so that the integration contour in $\sigma$ is along the real line; the case with negative values for the $R$-charges can be obtained by analytic continuation, deforming the contour. 
The $S^{2}$ partition function reads
\begin{eqnarray}
Z_{k,N}^{S^2} &=& \frac{1}{k!}\sum_{\vec{m}\in\mathbb{Z}^{k}} \int_{\mathbb{R}^{k}} \prod_{s=1}^{k} \frac{\mathrm{d} (r\sigma_{s})}{2\pi} e^{-4 \pi i \xi r \sigma_{s}-i\theta m_{s}} 
Z_{\text{gauge}} Z_{IJ}\, Z_{\text{adj}}\label{sipp}
\end{eqnarray}
where
\begin{equation}
Z_{\text{gauge}} = \prod_{s<t}^{k}\left(\dfrac{m_{st}^{2}}{4} + r^2\sigma_{st}^{2}\right)
\end{equation}
and the one-loop determinants of the matter contributions are given by
\begin{eqnarray}
Z_{IJ} &=& \prod_{s=1}^{k}\prod_{j=1}^{N}\frac{\Gamma\left(-i r \sigma_{s}+i r a_{j}-\frac{m_{s}}{2}\right)}{\Gamma\left(1+i r \sigma_{s}-i r a_{j}-\frac{m_{s}}{2}\right)}
\frac{\Gamma\left(i r \sigma_{s}-i r \left(a_{j}-\epsilon\right)+\frac{m_{s}}{2}\right)}{\Gamma\left(1-i r \sigma_{s}+i r \left(a_{j}-\epsilon\right)+\frac{m_{s}}{2}\right)} \label{f}
\\
Z_{\text{adj}} &=& \prod_{s,t=1}^{k}\frac{\Gamma\left(1-i r \sigma_{st}-i r \epsilon-\frac{m_{st}}{2}\right)}{\Gamma\left(i r \sigma_{st}+i r \epsilon-\frac{m_{st}}{2}\right)}
\frac{\Gamma\left(-i r \sigma_{st}+i r \epsilon_{1}-\frac{m_{st}}{2}\right)}{\Gamma\left(1+i r \sigma_{st}-i r \epsilon_{1}-\frac{m_{st}}{2}\right)}
\frac{\Gamma\left(-i r \sigma_{st}+i r \epsilon_{2}-\frac{m_{st}}{2}\right)}{\Gamma\left(1+i r \sigma_{st}-i r \epsilon_{2}-\frac{m_{st}}{2}\right)}\nonumber
\end{eqnarray}
with $\epsilon=\E + \EE$, $\sigma_{st} = \sigma_s - \sigma_t$ and $m_{st} = m_s - m_t$.
$Z_{IJ}$ contains the contributions from the chirals in the fundamental and antifundamental $I,J$, while $Z_{\text{adj}}$
the ones corresponding to the adjoint chirals $\chi,B_1,B_2$. 
The partition function (\ref{sipp}) is the central character of this paper and we will refer to it as the stringy instanton partition function.
Before closing this section, since this will play a r\^ole later in the paper, let us comment on the  
renormalization scheme used to define
the infinite products in the 1-loop determinant in the computation of the spherical partition function. In \cite{BC,gomisetal}
the $\zeta$-function renormalization scheme is chosen. Indeed this is a reference one, while others can be obtained 
by a shift in the finite part of the resulting effective action. 
These determinants appear in the form of ratios of Gamma-functions.
The ambiguity amounts to shift the Euler-Mascheroni constant $\gamma$ appearing in the 
Weierstrass form of the Gamma-function 
\beq
\frac{1}{\Gamma(x)}=x e^{\gamma x} \prod_{n=1}^\infty\left(1+\frac{x}{n}\right)e^{-\frac{x}{n}}
\label{gamma}
\eeq
with a finite function of the parameters. Due to supersymmetry, this function has to be encoded in terms of a holomorphic function $f(z)$, namely 
$\gamma\to {\rm Re}f(z)$. A more detailed discussion on this point will be performed in Sec. 3.

\subsection{Reduction to the Nekrasov partition function}

A first expected property of $Z^{S^2}_{k,N}$ is its reduction to the Nekrasov partition function in the limit of 
zero radius of the blown-up sphere.
Because of that, in (\ref{sipp}) we kept explicit the expression on the radius $r$.
It can easily be shown that 
in the limit $r \rightarrow 0$ our spherical partition function reduces to the integral representation of the 
instanton part of the Nekrasov partition function 
$Z_{N} = \sum_{k} \Lambda^{2 N k} Z_{k,N}^{\text{Nek}} $, where $Z_{k,N}^{\text{Nek}} $ is given by 
\begin{equation}
Z_{k,N}^{\text{Nek}} = \frac{1}{k!}\frac{\epsilon^{k}}{(2 \pi i \epsilon_{1} \epsilon_{2})^k} \oint 
\prod_{s=1}^k \dfrac{d\sigma_{s}}{P(\sigma_{s}) P(\sigma_{s}+\epsilon)}  \prod_{s<t}^k \dfrac{\sigma_{st}^2 
(\sigma_{st}^2 - \epsilon^{2} )}{(\sigma_{st}^2 - \epsilon_{1}^2 )(\sigma_{st}^2 - \epsilon_{2}^2 )}
\label{nek}
\end{equation} 
with $P(\sigma_{s})  = \prod_{j=1}^N (\sigma_{s}-a_{j})$ and $\Lambda$ the RGE invariant scale. \\
\noindent In order to prove this, let's start by considering \eqref{f}; because of the identity $\Gamma(z) = \Gamma(1+z)/z$, $Z_{IJ}$ and $Z_{\text{gauge}}Z_{\text{adj}}$ can be rewritten as
\begin{equation}
\begin{split}
Z_{IJ} \, = &\, \prod_{s=1}^{k}\prod_{j=1}^{N} \frac{1}{(r\sigma_s - r a_j -i \frac{m_s}{2})(r\sigma_s - r a_j + r \epsilon - i \frac{m_s}{2})}\\
&\prod_{s=1}^{k}\prod_{j=1}^{N}\frac{\Gamma\left(1-i r \sigma_{s}+i r a_{j}-\frac{m_{s}}{2}\right)}{\Gamma\left(1+i r \sigma_{s}-i r a_{j}-\frac{m_{s}}{2}\right)}
\frac{\Gamma\left(1+i r \sigma_{s}-i r \left(a_{j}-\epsilon\right)+\frac{m_{s}}{2}\right)}{\Gamma\left(1-i r\sigma_{s}+i r \left(a_{j}-\epsilon\right)+\frac{m_{s}}{2}\right)} \label{g1}
\end{split}
\end{equation} 
\begin{equation}
\begin{split}
& Z_{\text{gauge}} Z_{\text{adj}} \, = \, 
\prod_{s<t}^{k}\frac{\left(r\sigma_{st} + i \frac{m_{st}}{2} \right)\left(r\sigma_{st} - i \frac{m_{st}}{2} \right)\left(r\sigma_{st} + r \epsilon + i \frac{m_{st}}{2} \right)\left(r\sigma_{st} - r \epsilon + i \frac{m_{st}}{2} \right)}{\left(r\sigma_{st} - r \epsilon_1 - i \frac{m_{st}}{2} \right)\left(r\sigma_{st} + r \epsilon_1 - i \frac{m_{st}}{2} \right)\left(r\sigma_{st} - r \epsilon_2 - i \frac{m_{st}}{2} \right)\left(r\sigma_{st} + r \epsilon_2 - i \frac{m_{st}}{2} \right)}\\
& \left( \dfrac{\epsilon}{i r \epsilon_1 \epsilon_2} \right)^k \prod_{s\neq t}^{k}\frac{\Gamma\left(1-i r \sigma_{st}-i r \epsilon-\frac{m_{st}}{2}\right)}{\Gamma\left(1+i r \sigma_{st}+i r \epsilon-\frac{m_{st}}{2}\right)}
\frac{\Gamma\left(1-i r \sigma_{st}+i r \epsilon_{1}-\frac{m_{st}}{2}\right)}{\Gamma\left(1+i r \sigma_{st}-i r \epsilon_{1}-\frac{m_{st}}{2}\right)}
\frac{\Gamma\left(1-i r \sigma_{st}+i r \epsilon_{2}-\frac{m_{st}}{2}\right)}{\Gamma\left(1+i r \sigma_{st}-i r \epsilon_{2}-\frac{m_{st}}{2}\right)} \label{g2}
\end{split}
\end{equation} 
The lowest term in the expansion around $r = 0$ of \eqref{g1} comes from the $\vec{m} = \vec{0}$ sector, and is given by
\begin{equation}
\dfrac{1}{r^{2kN}}\prod_{s=1}^{k}\prod_{j=1}^{N} \frac{1}{(\sigma_s - a_j)(\sigma_s - a_j + \epsilon)}
\end{equation}
On the other hand, \eqref{g2} starts as
\begin{equation}
\left( \dfrac{\epsilon}{i r \epsilon_1 \epsilon_2} \right)^k (f(\vec{m}) + o (r))
\end{equation}
with $f(\vec{m})$ ratio of Gamma functions independent on $r$. With this, we can conclude that the first term in the expansion originates from the $\vec{m} = \vec{0}$ contribution, and \eqref{sipp} reduces to \eqref{nek}, with $\Lambda = qr^{-1}$ with $q$ being the classical instanton action contribution.

\subsection{Classification of the poles}

The explicit evaluation of the partition function \eqref{sipp} given above passes by the classification of the poles in the integrand. We now show that these are classified by Young tableaux, just like for the Nekrasov partition 
function \cite{Deng}. More precisely, we find a tower of poles for each box of the Young tableaux labelling the tower of Kaluza-Klein modes due to the string corrections.

The geometric phase of the GLSM is encoded in the choice of the contour of integration of \eqref{sipp}, which implements
the suitable stability condition for the hyper-K\"ahler quotient. In our case the ADHM phase corresponds to take 
$\xi>0$ and this imposes to close the contour integral in the lower half plane. Following the discussion of \cite{BC}, let us summarize the possible poles and zeros of the integrand ($n\geqslant 0$):
\begin{center}
\begin{tabular}{c|c|c}
& poles ($\sigma^{(p)}$) & zeros ($\sigma^{(z)}$) \\ 
\hline $I$   & $\sigma^{(p)}_s = a_j - \frac{i}{r}(n + \frac{\vert m_s \vert}{2})$ & $\sigma_s^{(z)} = a_j + \frac{i}{r}( 1+ n + \frac{\vert m_s \vert}{2})$ \\ 
\hline $J$   & $\sigma_s^{(p)} = a_j - \epsilon + \frac{i}{r}(n + \frac{\vert m_s \vert}{2})$ & $\sigma_s^{(z)} = a_j - \epsilon - \frac{i}{r}( 1+ n + \frac{\vert m_s \vert}{2})$ \\ 
\hline $\chi$ & $\sigma_{st}^{(p)} = - \epsilon - \frac{i}{r}(1 + n + \frac{\vert m_{st} \vert}{2})$ & $\sigma^{(z)}_{st} = - \epsilon + \frac{i}{r}(n + \frac{\vert m_{st} \vert}{2})$ \\ 
\hline $B_1$  & $\sigma^{(p)}_{st} = \epsilon_1 - \frac{i}{r}(n + \frac{\vert m_{st} \vert}{2})$ & $\sigma_{st}^{(z)} = \epsilon_1 + \frac{i}{r}(1+ n + \frac{\vert m_{st} \vert}{2})$ \\ 
\hline $B_2$  & $\sigma_{st}^{(p)} = \epsilon_2 - \frac{i}{r}(n + \frac{\vert m_{st} \vert}{2})$ & $\sigma_{st}^{(z)} = \epsilon_2 + \frac{i}{r}(1+n + \frac{\vert m_{st} \vert}{2})$ \\ 
\hline
\end{tabular} 
\end{center}
Poles from $J$ do not contribute, being in the upper half plane. Consider now a pole for $I$, say $\sigma_1^{(p)}$; the next pole $\sigma_2^{(p)}$ can arise from $I, B_1$ or $B_2$, but not from $\chi$, because in this case it would be cancelled by a zero from $J$. Moreover, if it comes from $I$, $\sigma_2^{(p)}$ should correspond to a twisted mass $a_j$ different from the one for $\sigma_1^{(p)}$, or the partition function would vanish (as explained in full detail in \cite{BC}). In the case $\sigma_2^{(p)}$ comes from $B_1$, consider $\sigma_3^{(p)}$: again, this can be a pole from $I, B_1$ or $B_2$, but not from $\chi$, or it would be cancelled by a zero of $B_2$. This reasoning takes into account all the possibilities, so we can conclude that the poles are classified by $N$ Young tableaux $\{\vec{Y}\}_k=\left(Y_1,\ldots,Y_N\right)$ such that $\sum_{j=1}^N |Y_j|=k$, which describe coloured partitions of the instanton number $k$. These are the same as the ones used in the pole classification of the Nekrasov partition function, with the difference that to every box is associated not just a pole, but an infinite tower of poles, labelled by a positive integer $n$; i.e., we are dealing with three-dimensional Young tableaux.

These towers of poles can be dealt with by rewriting near each pole 
\begin{equation}
\sigma_{s} = - \frac{i}{r}\left(n_s + \frac{\vert m_s \vert}{2}\right) + i \lambda_s
\end{equation} 
In this way we resum the contributions coming from the ``third direction'' of the Young tableaux, and the poles for $\lambda_s$ are now given in terms of usual two-dimensional partitions.
As we will discuss later, this procedure allows for a clearer geometrical interpretation
of the spherical partition function.
Defining $z = e^{-2\pi \xi+i\theta}$ and $d_s = n_s +\frac{m_s + \vert m_s \vert}{2}$, $\tilde{d}_s = d_s - m_s$ so that $\sum_{m_s\in \mathbb{Z}}\sum_{n_s\geqslant 0} = \sum_{\tilde{d}_s\geqslant 0}\sum_{d_s\geqslant 0}$ we obtain the following expression:
\begin{equation}
Z_{k,N}^{S^2} = 
\dfrac{1}{k!} \oint \prod_{s=1}^k \dfrac{d (r \lambda_{s})}{2\pi i} (z \bar{z})^{-r \lambda_{s}} Z_{\text{1l}} Z_{\text{v}} Z_{\text{av}} \label{int}
\end{equation}
where\footnote{Here and in the following, we will always be shifting $\theta \rightarrow \theta + (k-1)\pi$. This is needed in the non-abelian case in order to match $Z_{\text{v}}$ with the Givental $I$-functions known in the mathematical literature: we have in mind Grassmannians, flag manifolds \cite{BSTV}
and the Hilbert scheme of points in Sec.3.2 later on.}
\begin{eqnarray}
Z_{\text{1l}} &=& \left(\dfrac{\Gamma(1-i r \epsilon)\Gamma(i r \epsilon_{1})\Gamma(i r \epsilon_{2})}{\Gamma(i r \epsilon)\Gamma(1-i r \epsilon_{1})\Gamma(1-i r \epsilon_{2})}\right)^k \prod_{s=1}^k \prod_{j=1}^N \dfrac{\Gamma(r \lambda_{s}+i r a_{j})\Gamma(-r \lambda_{s}-i r a_{j}+ i r \epsilon)}{\Gamma(1 - r \lambda_{s}-i r a_{j})\Gamma(1+r \lambda_{s}+i r a_{j}- i r \epsilon)}\nonumber\\
&&\prod_{s\neq t}^k (r\lambda_{s}-r\lambda_{t})\dfrac{\Gamma(1+ r \lambda_{s}- r\lambda_{t}-i r \epsilon)\Gamma( r \lambda_{s}- r\lambda_{t} +i r \epsilon_{1})\Gamma( r \lambda_{s}- r\lambda_{t}+i r \epsilon_{2})}{\Gamma(-r\lambda_{s} + r \lambda_{t} + i r \epsilon)\Gamma(1-r\lambda_{s} + r \lambda_{t}-i r \epsilon_{1})\Gamma(1-r\lambda_{s} + r \lambda_{t}-i r \epsilon_{2})}\nonumber\\\label{1l}
\end{eqnarray}
\begin{eqnarray}
Z_{\text{v}} &=& \sum_{\tilde{d}_1,\ldots , \tilde{d}_k \,\geq \,0} ((-1)^N z)^{\tilde{d}_1+ \ldots +\tilde{d}_k}  \prod_{s=1}^k \prod_{j=1}^N \dfrac{(-r \lambda_{s}-i r a_{j}+ i r \epsilon)_{\tilde{d}_s}}{(1-r \lambda_{s}-i r a_{j})_{\tilde{d}_s}}
\prod_{s<t}^k \dfrac{\tilde{d}_t - \tilde{d}_s - r \lambda_{t} + r \lambda_{s}}{- r \lambda_{t} + r \lambda_{s}}\nonumber\\
&&\dfrac{(1+ r \lambda_{s}- r\lambda_{t}-i r \epsilon)_{\tilde{d}_t - \tilde{d}_s}}{( r \lambda_{s}- r\lambda_{t}+i r \epsilon)_{\tilde{d}_t - \tilde{d}_s}} 
\dfrac{( r \lambda_{s}- r\lambda_{t}+i r \epsilon_{1})_{\tilde{d}_t - \tilde{d}_s}}{(1+ r \lambda_{s}- r\lambda_{t}-i r \epsilon_{1})_{\tilde{d}_t - \tilde{d}_s}}
\dfrac{(r \lambda_{s}- r\lambda_{t}+i r \epsilon_{2})_{\tilde{d}_t - \tilde{d}_s}}{(1+ r \lambda_{s}- r\lambda_{t}-i r \epsilon_{2})_{\tilde{d}_t - \tilde{d}_s}}\nonumber\\ \label{v}
\end{eqnarray}
\begin{eqnarray}
Z_{\text{av}} &=& \sum_{d_1,\ldots , d_k \,\geq \,0} ((-1)^N \bar{z})^{d_1+ \ldots +d_k}  \prod_{s=1}^k \prod_{j=1}^N \dfrac{(-r \lambda_{s}-i r a_{j}+ i r \epsilon)_{d_s}}{(1-r \lambda_{s}-i r a_{j})_{d_s}}
\prod_{s<t}^k \dfrac{d_t - d_s - r \lambda_{t} + r \lambda_{s}}{- r \lambda_{t} + r \lambda_{s}}\nonumber\\
&&\dfrac{(1+ r \lambda_{s}- r\lambda_{t}-i r \epsilon)_{d_t - d_s}}{( r \lambda_{s}- r\lambda_{t}+i r \epsilon)_{d_t - d_s}} 
\dfrac{( r \lambda_{s}- r\lambda_{t}+i r \epsilon_{1})_{d_t - d_s}}{(1+ r \lambda_{s}- r\lambda_{t}-i r \epsilon_{1})_{d_t - d_s}}
\dfrac{(r \lambda_{s}- r\lambda_{t}+i r \epsilon_{2})_{d_t - d_s}}{(1+ r \lambda_{s}- r\lambda_{t}-i r \epsilon_{2})_{d_t - d_s}}\nonumber\\\label{av}
\end{eqnarray}
The Pochhammer symbol $(a)_d$ is defined as
\begin{equation}
(a)_d = \left\{ 
\begin{array}{cc}
\prod_{i=0}^{d-1} (a+i) & \,\,\text{for}\,\, d>0\\
1 & \,\,\text{for}\,\, d=0\\
\prod_{i=1}^{\vert d \vert} \dfrac{1}{a-i} & \,\,\text{for}\,\, d<0
\end{array}
\right.
\end{equation}
Notice that this definition implies the identity
\begin{equation}
(a)_{-d} = \dfrac{(-1)^d }{(1-a)_d} \label{poch}
\end{equation}
We observe that the $\frac{1}{k!}$ in \eqref{int} is cancelled by the $k!$ possible orderings of the $\lambda$s, so in the rest of this paper we will always choose an ordering and remove the factorial.
 
Let us remark that $Z_v$ appearing in \eqref{v} is the vortex partition function of the GLSM on equivariant $\mathbb{R}^2$ with equivariant parameter $\hbar=1/r$.
This was originally computed in \cite{shadchin} and recently discussed in the context of AGT correspondence in \cite{DGH,BTZ,gomisetal}.\\

As a final comment, let us consider the interesting limit $\epsilon_1 \rightarrow - \epsilon_2$, which implies $\epsilon \rightarrow 0$. In this limit we can show that all the world-sheet instanton corrections to $Z^{S^2}_{k,N}$ vanish and this is in agreement with the results of \cite{MO} about equivariant Gromov-Witten invariants of the ADHM moduli space. 

First of all, consider \eqref{1l}. The prefactor gives the usual coefficient $(\frac{\epsilon}{i\epsilon_1 \epsilon_2})^k$, while the Gamma functions simplify drastically, and we recover \eqref{nek} with $\epsilon$ small,  where the classical factor $(z\bar{z})^{-r \lambda_{r}}$ plays the r\^ole of the usual regulator in the contour integral representation of the Nekrasov partition function.
Let us now turn to \eqref{v}; for every Young tableau, we have $Z_{\text{v}} = 1 + o(\epsilon)$, where the 1 comes from the sector $\tilde{d}_s = 0$. 
Indeed one can show by explicit computation on the Young tableaux that for any $\tilde{d}_s \ne 0$ $Z_v$ gets a positive power of $\epsilon$ and 
therefore does not contribute in the $\epsilon\to 0$ limit.

To clarify this point, let us consider a few examples. We will restrict to $N=1$ for the sake of simplicity.
\begin{itemize}
\item The easiest tableau is $(\tiny\yng(1))$; in this case $\lambda = - i a$ and
\begin{equation}
Z_{\text{v}} \;=\; \sum_{\tilde{d}\,\geq \,0} (- z)^{\tilde{d}} \dfrac{( i r \epsilon)_{\tilde{d}}}{\tilde{d}!}
\; = \; 1 + \sum_{\tilde{d}\,\geq \,1} (- z)^{\tilde{d}} \dfrac{( i r \epsilon)_{\tilde{d}}}{\tilde{d}!} \;=\; 1 + o(\epsilon)
\end{equation}
\item Next are the tableaux $(\tiny\yng(1,1))$ and $\left(\tiny\yng(2)\right)$. The expression of $Z_{\text{v}}$ for $(\tiny\yng(1,1))$ is given in \eqref{ff}; there you can easily see from the two Pochhammers $(i r \epsilon)_{\tilde{d}}$ at the numerator that the limit $\epsilon \rightarrow 0$ forces the $\tilde{d}_s$ to be zero, leaving $Z_{\text{v}} = 1 + o(\epsilon)$; similarly for $\left(\tiny\yng(2)\right)$. 
\item The tableaux for $k = 3$ work as before. A more complicated case is $(\tiny\yng(2,2))$.
One should first consider the Pochhammers of type $(i r \epsilon)_{\tilde{d}}$ and $(1)_{\tilde{d}}$; in this case, we have
\begin{equation}
(i r \epsilon)_{\tilde{d}_1}(2i r \epsilon)_{\tilde{d}_4} \dfrac{(i r \epsilon)_{\tilde{d}_2 - \tilde{d}_1}(i r \epsilon)_{\tilde{d}_3 - \tilde{d}_1}(i r \epsilon)_{\tilde{d}_4 - \tilde{d}_2}(i r \epsilon)_{\tilde{d}_4 - \tilde{d}_3}(1)_{\tilde{d}_4 - \tilde{d}_1}}{(2i r \epsilon)_{\tilde{d}_4 - \tilde{d}_1}
(1)_{\tilde{d}_2 - \tilde{d}_1}(1)_{\tilde{d}_3 - \tilde{d}_1}(1)_{\tilde{d}_4 - \tilde{d}_2}(1)_{\tilde{d}_4 - \tilde{d}_3}} \dfrac{\tilde{d}_4 - \tilde{d}_1 + i r \epsilon}{i r \epsilon}
\end{equation}
Then one can easily see that this combination either starts with something which is of order $\epsilon$ or higher,
or is zero (i.e the contribution $\tilde{d}_1 \neq 0$, $\tilde{d}_2 = \tilde{d}_3 = \tilde{d}_4 = 0$), unless 
every $\tilde{d}_s = 0$, in which case we get 1.
\end{itemize}
These examples contain all the possible issues that can arise in the general case. 


\section{Equivariant Gromov-Witten invariants of the instanton moduli space}
\label{sec:GW}

We now turn to discuss the exact partition function \eqref{int} of the D1-D5 system on the resolved $A_1$
singularity. As discussed in the previous section, this contains a tower of non-perturbative 
corrections 
to the prepotential of the four-dimensional gauge theory
corresponding to the 
effective
world-sheet instantons contributions. We will show in this section that these corrections compute the Gromov-Witten invariants
and gravitational descendants of the ADHM moduli space. 
It has been argued in \cite{morrison} and shown in \cite{gomislee} that the spherical partition
function computes the vacuum amplitude of the non-linear $\sigma$-model (NLSM) in the infrared 
\beq
\langle\bar 0 | 0 \rangle= e^{-K}
\label{00}
\eeq
where $K$ is the quantum K\"ahler potential of the target space $X$. Let us rewrite the above vacuum amplitude
in a way which is more suitable for our purposes.
Following \cite{BCOV,Dubrovin}, let us introduce the flat sections spanning the vacuum bundle
satisfying
\beq
\left(\hbar D_a \delta_b^c + C_{ab}^c\right)V_c=0.
\label{1}\eeq
where $D_a$ is the covariant derivative on the vacuum line bundle and $C_{ab}^c$ are the
coefficients of the OPE in the chiral ring of observables $\phi_a\phi_b =C_{ab}^c \phi_c$. 
The observables $\{\phi_a\}$ provide a basis for the vector space
of chiral ring operators $H^0(X)\oplus H^2(X)$ with $a=0,1,\ldots,b^2(X)$, 
$\phi_0$ being the identity operator.
The parameter $\hbar$ is the spectral parameter of the Gauss-Manin connection. 
Specifying the case $b=0$ in (\ref{1}), we find that 
$$
V_a=-\hbar D_aV_0
$$
this means that the flat sections are all generated by the fundamental solution ${\cal J}:=V_0$
of the equation
\beq
\left(\hbar D_a D_b + C_{ab}^c D_c\right){\cal J}=0
\label{2}\eeq
The metric on the vacuum bundle is given by a symplectic pairing of the flat sections
$$g_{\bar a b}= \langle \bar a|b\rangle = V_{\bar a}^t E V_b$$
and in particular
the vacuum-vacuum amplitude, that is
the spherical partition function, can be written as the symplectic pairing
\beq
\langle\bar 0 | 0 \rangle = {\cal J}^t E {\cal J}
\label{vac}
\eeq
for a suitable symplectic form $E$ \cite{BCOV} that will be specified later for our case.
We remark that since the ADHM moduli space is a non-compact holomorphic symplectic manifold,
the world-sheet instanton corrections are non-trivial only in presence of a non-vanishing $\Omega$-background.  
From the mathematical viewpoint, this amounts to work in the context of equivariant cohomology of the target space $H^\bullet_T(X)$
where $T$ is the torus acting on $X$ \cite{coates-giv}. For example, for local ${\mathbb P}^1$ geometries this has interesting connections with integrable systems 
\cite{braini}.

We point out that there is a natural correspondence of the results of supersymmetric localization
with the formalism developed by Givental for the computation of the flat section ${\cal J}$.
Indeed, as we have discussed in Sec.2,
the computation of the spherical partition function via localization 
makes use of a supersymmetric charge which closes on a $U(1)$ isometry
of the sphere.
This is precisely the setting considered by Givental in \cite{egw} to describe $S^1$-equivariant Gromov-Witten
invariants. Indeed, in this approach one considers
holomorphic maps which are equivariant with respect to the maximal torus of the sphere automorphisms 
$S^1\subset PSL(2,\mathbb{C})$. This is to be identified with the $U(1)$ isometry on which
the supersymmetry algebra closes. As a consequence,  
the equivariant parameter $\hbar$ of Givental's $S^1$ action gets identified with the one of the 
vortex partition functions arising in the localization of the spherical partition function.

Since Givental's formalism plays a major r\^ole in the subsequent discussion, let us first describe it briefly,
see \cite{coates-giv} for details.
Givental's small ${\cal J}$-function is given by the $H_T^0(X)\oplus H^2_T(X)$ valued generating function
\beq
{\cal J}_X(\tau,\hbar) = e^{\tau/\hbar}\left(1 + \sum_d Q^d e^{d\tau} \left\langle \frac{\phi^a}{\hbar(\hbar - \psi_1)}\right\rangle_{X_{0,1,d}}\phi_a\right)
\label{J}
\eeq
where $\tau=\tau^a\phi_a$, $\ \ \psi_{1}$ is the gravitational descendant insertion at the marked point and the sigma model expectation value
localizes on the moduli space $X_{0,1,d}$ of holomorphic maps of degree $d\in\mathbb{N}_{>0}$ from the sphere with one marked point to the target space $X$.
The world-sheet instanton corrections are labelled by
the parameter $Q^d = \prod_{i=1}^{b_2(X)} Q_i^{d_i}$ with $Q_i = e^{- t^i}$, $t^i$ being the complexified K\"ahler parameters.
 
Givental has shown how to reconstruct the ${\cal J}$-function from a set of oscillatory integrals, the so called ``${\cal I}$-functions''
which are generating functions of hypergeometric type in the variables $\hbar$ and $Q$. 
We observe that Givental's formalism has been developed originally for abelian quotients, more precisely for complete intersections in quasi-projective toric varieties.
In this case, the ${\cal I}$ function is the generating function of solutions of the Picard-Fuchs equations for the mirror manifold $\check X$ of $X$ 
and as such can be expressed in terms of periods on $\check X$.

This formalism has been extended to non-abelian GLSM in \cite{ciocan1,ciocan2}. The Gromov-Witten invariants for the non-abelian quotient $M//G$
are conjectured to be expressible in terms of the ones of the corresponding abelian quotient $M//T$, $T$ being the maximal torus of $G$,
twisted by the Euler class of a vector bundle over it. The corresponding ${\cal I}$-function is obtained from the one associated to the abelian quotients
multiplied by a suitable factor depending on the Chern roots of the vector bundle.  
The first example of this kind was the quantum cohomology of the
Grassmanian discussed in \cite{HV}. This was rigorously proved and extended to flag manifolds in \cite{ciocan1}.
As we will see, our results give evidence of the above conjecture in full generality, though a rigorous mathematical proof
of this result is not available at the moment, see \cite{BSTV} for a more detailed discussion.\footnote{A related issue concerning the equivalence of symplectic quotients
and GIT quotients via the analysis of vortex moduli space has been also discussed in \cite{amico}.}  

In order to calculate the equivariant Gromov-Witten invariants from the above functions, one has to consider their asymptotic
expansion in $\hbar$. It is clear from \eqref{J} that up to the exponential prefactor, the ${\cal J}$ function expands as
\beq
1 + \frac{J^{(2)}}{\hbar^2} +  \frac{J^{(3)}}{\hbar^3} + \ldots
\label{Jexp}
\eeq
such that each coefficient is a cohomology-valued formal power series in the $Q$-variables. 
We observe that the coefficient of the $\hbar^{-2}$ term in the expansion is directly related to the Gromov-Witten
prepotential ${\cal F}$. Indeed from \eqref{J} we deduce that $J^{(2)a} = \eta^{ab}\partial_b{\cal F}$
where $\eta^{ab}$ is the inverse topological metric. Higher order terms in \eqref{Jexp} are related to gravitational
descendant insertions.

The analogous expansion for ${\cal I}_X(q,\hbar)$ reads 
\beq
I^{(0)} + \frac{I^{(1)}}{\hbar}
+  \frac{I^{(2)}}{\hbar^2} + \ldots
\label{Iexp}
\eeq 
and the coefficients $I^{(0)}$, $I^{(1)}$ provide the change of variables which transforms ${\cal I}$ into ${\cal J}$ defining the equivariant mirror map. 
If ${I}^{(1)}=0$ the equivariant mirror map is trivial and the two functions coincide. \footnote{Usually, one splits $I^{(1)} = \sum_s p_s g^s(z) + \sum_i\widetilde{p}_i h^i(z)$, with $p_s$ cohomology generators and $\widetilde{p}_i$ equivariant parameters of $H^2_T(X)$. The functions $\mathcal{I}$ and $\mathcal{J}$ are related by $\mathcal{J}(\hbar, q) = e^{-f_0(z)/\hbar} e^{- \sum_i \widetilde{p}_i h^i (z)/\hbar} \mathcal{I}(\hbar, z(q))$; comparing with our notation, we identify $f_0(z)/\hbar = \ln I^{(0)}$. The mirror map is given by (in the simple example with just one $p$ and $\widetilde{p}$) $q = \ln z + \frac{g(z)}{I^{(0)}(z)}$, because of a normalization for $\mathcal{I}$ slightly different from the standard one; in our case, $g(z) = 0$ since $p = 0$ in the fully equivariant case. The factor $e^{- \sum_i \widetilde{p}_i h^i (z)/\hbar}$ is the so-called equivariant mirror map, and has to be identified with what we call ${\cal N}_{\rm v}$.}

We are now ready to state the dictionary between Givental's formalism and the spherical partition function
\beq
{Z}^{S^2} = \oint 
d\lambda Z_{1l} \left(z^{-r|\lambda|} Z_{\rm v}\right) \left(\bar z^{-r|\lambda|} Z_{\rm av}\right)
\label{ciao} 
\eeq
with $d\lambda=\prod_{\alpha=1}^{\rm rank}d\lambda_\alpha$ and $|\lambda|=\sum_\alpha \lambda_\alpha$.
Our claim is that $Z_{\rm v}$ is to be identified with Givental ${\cal I}$ function 
upon identifying the vortex counting parameter $z$ with $Q$, $\lambda_\alpha$ with the generators of the
equivariant cohomology and $r=1/\hbar$. To extract the Gromov-Witten invariants from the spherical partition function
one has then to implement the procedure outlined above to compute the ${\cal J}$ function.
This is obtained by choosing a suitable normalization factor ${\cal N}_{\rm v}$ such that the resulting vortex partition function
has the same expansion as \eqref{Jexp}. 
From the viewpoint of the quantum K\"ahler potential, this normalization 
fixes the proper K\"ahler frame in which \eqref{vac} holds, thus
\beq
{\cal J}^t E {\cal J} = \frac{Z^{S^2}}{|{\cal N}_{tot}|^2}
\label{N}
\eeq
and ${\cal J}={\cal I}/{\cal N}_{\rm v}$. Notice that the normalization factor ${\cal N}_{\rm v}$ is $\lambda$ independent and that the symplectic pairing
$E$ is provided by the contour integral in the $\lambda$s with measure given by the one-loop partition function $Z_{1l}$ as appearing in \eqref{ciao}. 
Actually, also the symplectic pairing has to be properly normalized to give the classical equivariant intersection of the target space. Henceforth
the overall normalization ${\cal N}_{tot}= {\cal N}_{1l} {\cal N}_{\rm v}$ appears in \eqref{N}.
This amounts to a suitable choice of the renormalization scheme for the one-loop determinants appearing in $Z^{S^2}$ which ensures the spherical partition
function to reproduce the correct classical (equivariant) intersection pairing of the cohomology of the target space in the $r\to 0$ limit. This has to be found in a
case by case analysis and will be specified further in the 
examples discussed in the following subsections.

From the above discussion we deduce that the spherical partition function of the D1-D5 GLSM provide
conjectural formulae for Givental's ${\cal I}$ and ${\cal J}$-functions of the ADHM instanton moduli space
as follows
\begin{eqnarray}
{\cal I}_{k,N}&=& \sum_{d_1,\ldots , d_k \,\geq \,0} ((-1)^N z)^{d_1+ \ldots +d_k}  \prod_{s=1}^k \prod_{j=1}^N \dfrac{(-r \lambda_{s}-i r a_{j}+ i r \epsilon)_{d_s}}{(1-r \lambda_{s}-i r a_{j})_{d_s}}
\prod_{s<t}^k \dfrac{d_t - d_s - r \lambda_{t} + r \lambda_{s}}{- r \lambda_{t} + r \lambda_{s}}\nonumber\\
&&\dfrac{(1+ r \lambda_{s}- r\lambda_{t}-i r \epsilon)_{d_t - d_s}}{( r \lambda_{s}- r\lambda_{t}+i r \epsilon)_{d_t - d_s}} 
\dfrac{( r \lambda_{s}- r\lambda_{t}+i r \epsilon_{1})_{d_t - d_s}}{(1+ r \lambda_{s}- r\lambda_{t}-i r \epsilon_{1})_{d_t - d_s}}
\dfrac{(r \lambda_{s}- r\lambda_{t}+i r \epsilon_{2})_{d_t - d_s}}{(1+ r \lambda_{s}- r\lambda_{t}-i r \epsilon_{2})_{d_t - d_s}}\nonumber\\
\label{I-inst}\end{eqnarray}
where $\lambda_s$ are the Chern roots of the tautological bundle of the ADHM moduli space.
 
From the above expression we find that the asymptotic behaviour in $\hbar$ is 
\beq
{\cal I}_{k,N}= 1 + \frac{I^{(N)}}{\hbar^N} +\ldots
\eeq
Therefore, $I^{(0)} = 1$ for every $k, N$, while $I^{(1)} = 0$ when $N>1$; this implies that the equivariant mirror map is trivial, namely ${\cal I}_{k,N}={\cal J}_{k,N}$, for $N>1$. The $N=1$ case will be discussed in detail in the following subsection.
The structure of \eqref{I-inst} supports the abelian/non-abelian correspondence conjecture of \cite{ciocan}; indeed the first factor
in the first line corresponds to the abelian quotient by the Cartan torus $\left(\mathbb{C}^*\right)^k$ while the remaining factors
express the twisting due to the non-abelian nature of the quotient.

Finally, let us notice\footnote{We thank D.E. Diaconescu, A. Okounkov and D. Maulik for 
clarifying discussions on this issue.} 
that for GIT quotients, and in particular for Nakajima quiver varieties, the notion
of quasimaps 
and of the corresponding ${\cal I}$-function were introduced in
\cite{CKM}. We notice that our ${\cal I}_{k,N}$ as in \eqref{I-inst} should match 
the quasi-map ${\cal I}$-function and therefore, as a consequence of \cite{MO}, should 
compute the ${\cal J}$-function of the instanton moduli space. Let us underline
that the supersymmetric localization approach applies also to other classical groups
and then can be applied to study the quantum cohomology of general K\"ahler quotients.

\subsection{Cotangent bundle of the projective space}

As a first example, let us consider the case $\mathcal{M}_{1,N}\simeq \mathbb{C}^{2}\times T^{*}\mathbb{CP}^{N-1}$. The integrated spherical partition function has the form:
\begin{equation}
Z_{1,N} = \sum_{j=1}^{N} (z\bar{z})^{i r a_{j}} Z_{\text{1l}}^{(j)} Z_{\text{v}}^{(j)} Z_{\text{av}}^{(j)}
\end{equation}
The $j$-th contribution comes from the Young tableau $(\bullet \,,\,\ldots\,,\,{\tiny\yng(1)}\,,\,\ldots\,,\,\bullet)$, where the box is in the $j$-th position; this means we have to consider the pole $\lambda_{1} = - i a_{j}$. Explicitly:
\begin{eqnarray}
Z_{\text{1l}}^{(j)} &=& \frac{\Gamma\left(i r \epsilon_{1}\right)\Gamma\left(i r \epsilon_{2}\right)}{\Gamma\left(1-i r \epsilon_{1}\right)\Gamma\left(1-i r \epsilon_{2}\right)} \prod_{\substack{l=1 \\ l\neq j}}^{N} \frac{\Gamma\left(i r a_{lj}\right)\Gamma\left(-i r a_{lj}+i r \epsilon\right)}{\Gamma\left(1-i r a_{lj}\right)\Gamma\left(1+i r a_{lj}-i r \epsilon\right)}\nonumber\\
Z_{\text{v}}^{(j)} &=& {}_{N}F_{N-1}\left(\begin{array}{cc}\left\{i r \epsilon,\left(-i r a_{lj}+i r \epsilon\right)_{\substack{l=1\\l\neq j}}^{N}\right\}\\ \left\{\left(1-i r a_{lj}\right)_{\substack{l=1 \\ l\neq j}}^{N}\right\}\end{array};\left(-1\right)^{N}z \right)\nonumber\\
Z_{\text{av}}^{(j)} &=& {}_{N}F_{N-1}\left(\begin{array}{cc}\left\{i r \epsilon,\left(-i r a_{lj}+i r \epsilon\right)_{\substack{l=1\\l\neq j}}^{N}\right\}\\ \left\{\left(1-i r a_{lj}\right)_{\substack{l=1 \\ l\neq j}}^{N}\right\}\end{array};\left(-1\right)^{N}\bar{z} \right)
\end{eqnarray}
Let us consider in more detail the case $N=2$. In this case the instanton moduli space reduces to $\mathbb{C}^2\times T^*\mathbb{P}^1$ and is the same as the moduli space
of the Hilbert scheme of two points
 $\mathcal{M}_{1,2} \simeq \mathcal{M}_{2,1}$.
In order to match the equivariant actions on the two moduli spaces, we identify
\begin{equation}
a_1 = \epsilon_{1} + 2 a \;\;\;\;\;\;,\;\;\;\;\;\; a_2 = \epsilon_2 + 2a
\end{equation} 
so that $a_{12} = \epsilon_{1} - \epsilon_{2}$. Then we have
\begin{eqnarray}
Z_{1,2} = (z\bar{z})^{i r (2 a + \epsilon_{1})} Z_{\text{1l}}^{(1)} Z_{\text{v}}^{(1)} Z_{\text{av}}^{(1)} + 
(z\bar{z})^{i r (2 a + \epsilon_{2})} Z_{\text{1l}}^{(2)} Z_{\text{v}}^{(2)} Z_{\text{av}}^{(2)}\label{12} 
\end{eqnarray}
where
\begin{eqnarray}
Z^{(1)}_{\text{1l}} &=& \frac{\Gamma\left(i r \epsilon_{1}\right)\Gamma\left(i r \epsilon_{2}\right)}{\Gamma\left(1-i r \epsilon_{1}\right)\Gamma\left(1-i r \epsilon_{2}\right)}  \frac{\Gamma\left(-i r \epsilon_{1}+i r \epsilon_{2}\right)\Gamma\left(2 i r \epsilon_{1}\right)}{\Gamma\left(1+ i r \epsilon_{1}- i r \epsilon_{2}\right)\Gamma\left(1-2 i r \epsilon_{1}\right)}\nonumber\\
Z^{(1)}_{\text{v}} &=& {}_{2}F_{1}\left(\begin{array}{cc}\left\{i r \epsilon ,2 i r \epsilon_{1} \right\}\\ \left\{1+i r \epsilon_{1}-i r \epsilon_{2}\right\}\end{array}; z \right)\nonumber\\
Z^{(1)}_{\text{av}} &=& {}_{2}F_{1}\left(\begin{array}{cc}\left\{i r \epsilon ,2 i r \epsilon_{1} \right\}\\ \left\{1+i r \epsilon_{1}-i r \epsilon_{2}\right\}\end{array}; \bar{z} \right)
\end{eqnarray}
The other contribution is obtained by exchanging $\epsilon_{1} \longleftrightarrow \epsilon_{2}$. By identifying $Z^{(1)}_{\text{v}}$ as the Givental 
${\cal I}$-function, we expand it in $r = \frac{1}{\hslash}$ in order to find the equivariant mirror map; this gives
\begin{equation}
Z^{(1)}_{\text{v}} = 1 + o(r^2),
\end{equation}
which means there is no equivariant mirror map and ${\cal I} = {\cal J}$. The same applies to $Z^{(2)}_{\text{v}} $.

Therefore, the only normalization to be dealt with is the one of the symplectic pairing, namely $Z_{1{\rm l}}$. As discussed in Sec.2, there is a finite
term ambiguity related to the choice of the renormalization scheme of the one-loop determinants. In general the implementation of $\zeta$-function
regularization induces the presence of terms in the Euler-Mascheroni constant $\gamma$. In particular, as it follows immediately from \eqref{gamma}, this
happens if the sum of the arguments of the Gamma-functions in $Z_{1{\rm l}}$ is different from zero. 
In order to compensate these terms, we multiply by an appropriate ratio of Gamma functions which starts with $1$ in the $r$ expansion and makes the 
overall argument zero; this sets $\gamma$ to zero. 
Moreover, a renormalization scheme exists such that the expansion of the spherical partition function reproduces the classical intersection pairing
of equivariant cohomology \footnote{Similar arguments appeared also in \cite{manabe}.}. In order to get the correct result we then
multiply by a further factor of $z\bar z$ to the suitable power. 
  
Let us see how this works in our example. In \eqref{12}, $Z^{(1)}_{\text{1l}}$ and $Z^{(2)}_{\text{1l}}$ contain an excess of $4 i r (\epsilon_{1}+\epsilon_{2})$ in the argument of the Gamma functions ($2 i r (\epsilon_{1}+\epsilon_{2})$ from the numerator and another $2 i r (\epsilon_{1}+\epsilon_{2})$ from the denominator); this would imply 
\begin{equation}
Z^{(1)}_{\text{1l}} = -\dfrac{1}{2\E^2 \EE (\E - \EE) r^4 } + \dfrac{2 i \gamma \epsilon}{\E^2 \EE (\E - \EE) r^3 } + o(r^{-2})
\end{equation}
and similarly for $Z^{(2)}_{\text{1l}}$. To eliminate the Euler-Mascheroni constant, we normalize the partition function multiplying it by\footnote{The normalization here has been chosen having in mind the $\mathcal{M}_{2,1}$ case; see the next paragraph.}
\begin{equation}
(z \bar{z})^{-2 i r a} \left(\dfrac{\Gamma(1- i r \epsilon_{1})\Gamma(1- i r \epsilon_{2})}{\Gamma(1+ i r \epsilon_{1})\Gamma(1+ i r \epsilon_{2})}\right)^2 \label{norm}
\end{equation} 
so that now we have
\begin{equation}
\left(\dfrac{\Gamma(1- i r \epsilon_{1})\Gamma(1- i r \epsilon_{2})}{\Gamma(1+ i r \epsilon_{1})\Gamma(1+ i r \epsilon_{2})}\right)^2 Z^{(1)}_{\text{1l}} = -\dfrac{1}{2\E^2 \EE (\E - \EE) r^4 } + o(r^{-2})
\end{equation}
Expanding the normalized partition function in $r$ up to order $r^{-1}$, we obtain \footnote{Notice that the procedure outlined above does not fix a remnant dependence on the coefficient
of the $\zeta(3)$ term in $Z^{S^2}$. In fact, one can always multiply by a ratio of Gamma functions whose overall argument is zero; 
this will have an effect only on the $\zeta(3)$ coefficient. This ambiguity does not affect the calculation of the Gromov-Witten invariants.} 
\begin{eqnarray}
Z_{1,2}^{\text{norm}} &=& \dfrac{1}{r^2 \epsilon_1 \epsilon_2} \Big[ \dfrac{1}{2 r^2 \epsilon_{1}\epsilon_{2}} +\dfrac{1}{4}\ln^{2} (z \bar{z}) -i r(\epsilon_1 + \epsilon_2) \Big( -\dfrac{1}{12} \ln^{3} (z \bar{z}) - \ln (z \bar{z})(\text{Li}_2(z) + \text{Li}_2(\bar{z})) \nonumber\\
&+& 2 (\text{Li}_3(z) + \text{Li}_3(\bar{z})) + 3 \zeta(3) \Big)  \Big] \label{12kahler}
\end{eqnarray}
The first term in \eqref{12kahler} correctly reproduces the Nekrasov partition function of ${\cal M}_{1,2}$ as expected, while the other terms compute 
the $H^2_T(X)$ part of the genus zero Gromov-Witten potential in agreement with \cite{BG}. We remark that the quantum part of the Gromov-Witten potential
turns out to be linear in the equivariant parameter $\epsilon_1+\epsilon_2$ as inferred in Sect.2 from general arguments.
 
We can also compute it with the Givental formalism: expanding the $J$ function up to order $r^2$, one finds
\begin{equation}
J = 1 + r^2 (- \epsilon_1 \epsilon_2 - i (\epsilon_1 + \epsilon_2) \lambda_{1} + \lambda_{1}^2) \text{Li}_2(z) + o(r^3)
\end{equation}
and the coefficient of $-\lambda_{1}$ -- which is the cohomology generator -- at order $r^2$ will give the first $z$ derivative of the prepotential. 

  
\subsection{Hilbert scheme of points}

Let us now turn to the $\mathcal{M}_{k,1}$ case, which corresponds to the Hilbert scheme of $k$ points. 
This case was analysed in terms of Givental formalism in \cite{Fontanine}. It is easy to see that \eqref{I-inst} reduces for $N=1$
to their results.
As remarked after equation \eqref{I-inst} in the $N=1$ case there is a non-trivial equivariant mirror map to be implemented.
As we will discuss in a moment, this is done by defining the ${\cal J}$ function as
${\cal J} = (1+z)^{i r k \epsilon}{\cal I}$, which corresponds to invert the equivariant mirror map; in other words, we have to normalize the vortex part by multiplying it with $(1+z)^{i r k \epsilon}$, and similarly for the antivortex.
In the following we will describe in detail some examples and extract the relevant Gromov-Witten invariants for them. As we will see, these are in agreement
with the results of \cite{OP}. 

For $k=1$, the only Young tableau $({\tiny\yng(1)})$ corresponds to the pole $\lambda_{1} = -i a $. This case is simple enough to be written in a closed form; we find
\begin{equation}
Z^{S^2}_{1,1} = (z \bar{z})^{i r a} \dfrac{\Gamma(i r \epsilon_{1})\Gamma(i r \epsilon_{2})}{\Gamma(1-i r \epsilon_{1})\Gamma(1-i r \epsilon_{2})} (1+z)^{-i r \epsilon}(1+\bar{z})^{-i r \epsilon}\label{Z11}
\end{equation}
From this expression, it is clear that the Gromov-Witten invariants are vanishing. \\
\noindent Actually, we should multiply \eqref{Z11} by $(1+z)^{i r \epsilon}(1+\bar{z})^{i r \epsilon}$ in order to recover the ${\cal J}$-function. Instead of doing this, we propose to use $Z_{1,1}$ as a normalization for $Z_{k,1}$ as
\begin{equation}
Z_{k,1}^{\text{norm}} = \dfrac{Z^{S^2}_{k,1}}{(- r^2 \epsilon_{1}\epsilon_{2} Z^{S^2}_{1,1})^k}
\label{zio}
\end{equation} 
In this way, we go from ${\cal I}$ to ${\cal J}$ functions and at the same time we normalize the 1-loop factor in such a way to erase the Euler-Mascheroni constant. The factor $(- r^2 \epsilon_{1}\epsilon_{2})^k$ is to make the normalization factor to start with 1 in the $r$ expansion. In summary, we obtain
\begin{equation}
Z_{1,1}^{\text{norm}} = - \dfrac{1}{r^2 \E \EE}
\end{equation}
Let us make a comment on the above normalization procedure. From the general arguments discussed in the opening part of this Section we expect the normalization
to be independent on $\lambda$s. Moreover, from the field theory viewpoint, the normalization \eqref{zio} is natural since amounts to remove from the free energy
the contribution of $k$ free particles. On the other hand, this is non trivial at all from the explicit expression of the ${\cal I}$-function \eqref{I-inst}. Actually
a remarkable combinatorial identity proved in \cite{Fontanine} ensures that $e^{-I^{(1)}/\hbar}= (1+z)^{ik(\E+\EE)/\hbar}$ and then makes this procedure consistent.

Let us now turn to the $\mathcal{M}_{2,1}$ case. There are two contributions, $({\tiny\yng(1,1)})$ and $({\tiny\yng(2)})$, coming respectively from the poles $\lambda_{1} = -i a, \lambda_{2} = -i a -i \epsilon_{1}$ and $\lambda_{1} = -i a, \lambda_{2} = -i a -i \epsilon_{2}$. Notice once more that the  permutations of the $\lambda$'s are cancelled against the $\frac{1}{k!}$ in front of the partition function \eqref{sipp}. We thus have
\begin{eqnarray}
Z^{S^2}_{2,1} = (z\bar{z})^{i r (2 a + \epsilon_{1})} Z_{\text{1l}}^{(\text{col})} Z_{\text{v}}^{(\text{col})} Z_{\text{av}}^{(\text{col})} + 
(z\bar{z})^{i r (2 a + \epsilon_{2})} Z_{\text{1l}}^{(\text{row})} Z_{\text{v}}^{(\text{row})} Z_{\text{av}}^{(\text{row})}\label{21} 
\end{eqnarray}
where, explicitly,
\begin{equation}
\begin{split}
Z_{\text{1l}}^{(\text{col})} \,=&\, \dfrac{\Gamma(i r \epsilon_{1})\Gamma(i r \epsilon_{2})}{\Gamma(1-i r \epsilon_{1})\Gamma(1-i r \epsilon_{2})}  \dfrac{\Gamma(2i r \epsilon_{1})\Gamma(i r \epsilon_{2}-i r \epsilon_{1})}{\Gamma(1-2i r \epsilon_{1})\Gamma(1+i r \epsilon_{1}-i r \epsilon_{2})} \\
Z_{\text{v}}^{(\text{col})}\,=&\, \sum_{\tilde{d} \geqslant 0}(-z)^{\tilde{d}} \sum_{\tilde{d}_1 = 0}^{\tilde{d}/2}\dfrac{(1+i r \epsilon_{1})_{\tilde{d}-2 \tilde{d}_1}}{(i r \epsilon_{1})_{\tilde{d}-2 \tilde{d}_1}}\dfrac{(i r \epsilon)_{\tilde{d}_{1}}}{\tilde{d}_1 !} \dfrac{(i r \epsilon_{1}+i r \epsilon)_{\tilde{d}- \tilde{d}_1}}{(1+i r \epsilon_{1})_{\tilde{d}- \tilde{d}_1}}\\
&\dfrac{(2 i r \epsilon_{1})_{\tilde{d}-2 \tilde{d}_1}}{ (\tilde{d}-2 \tilde{d}_1) !} \dfrac{(1-i r \epsilon_{2})_{ \tilde{d}-2 \tilde{d}_1}}{(i r \epsilon_{1}+i r \epsilon)_{\tilde{d}-2 \tilde{d}_1}} \dfrac{(i r \epsilon)_{ \tilde{d}-2 \tilde{d}_1}}{(1+i r \epsilon_{1}-i r \epsilon_{2})_{ \tilde{d}-2 \tilde{d}_1}}\\
Z_{\text{av}}^{(\text{col})} \,=&\, \sum_{d \geqslant 0} (-\bar{z})^d \sum_{d_1 = 0}^{d/2} \dfrac{(1+i r \epsilon_{1})_{d-2 d_1}}{(i r \epsilon_{1})_{d-2 d_1}}\dfrac{(i r \epsilon)_{d_{1}}}{d_1 !} \dfrac{(i r \epsilon_{1}+i r \epsilon)_{d- d_1}}{(1+i r \epsilon_{1})_{d- d_1}}\\
&\dfrac{(2 i r \epsilon_{1})_{d-2 d_1}}{ (d-2 d_1) !} \dfrac{(1-i r \epsilon_{2})_{ d-2 d_1}}{(i r \epsilon_{1}+i r \epsilon)_{d-2 d_1}} \dfrac{(i r \epsilon)_{ d-2 d_1}}{(1+i r \epsilon_{1}-i r \epsilon_{2})_{ d-2 d_1}} \label{ff}
\end{split}
\end{equation}  
Here we defined $d = d_1 + d_2$ and changed the sums accordingly. The row contribution can be obtained from the column one by exchanging $\epsilon_{1} \longleftrightarrow \epsilon_{2}$. 
 We then have
\begin{equation}
Z_{\text{v}}^{(\text{col, row})} = 1 + 2 i r \epsilon   \text{Li}_1(-z) + o(r^2)
\end{equation}
Finally, 
we invert the equivariant mirror map by replacing 
\begin{eqnarray}
Z_{\text{v}}^{(\text{col, row})} \longrightarrow e^{-2 i r \epsilon  \text{Li}_1(-z)} Z_{\text{v}}^{(\text{col, row})} = (1+z)^{2 i r \epsilon}Z_{\text{v}}^{(\text{col, row})} \nonumber\\
Z_{\text{av}}^{(\text{col, row})} \longrightarrow e^{-2 i r \epsilon   \text{Li}_1(-\bar{z})} Z_{\text{av}}^{(\text{col, row})} = (1+\bar{z})^{2 i r \epsilon}Z_{\text{av}}^{(\text{col, row})}
\end{eqnarray}
Now we can prove the equivalence $\mathcal{M}_{1,2} \simeq \mathcal{M}_{2,1}$: by expanding in $z$, it can be shown that $Z^{(1)}_{\text{v}}(z) = (1+z)^{2 i r \epsilon}Z_{\text{v}}^{(\text{col})}(z)$ and similarly for the antivortex part; since $Z_{\text{1l}}^{(1)} =  Z_{\text{1l}}^{(\text{col})}$ we conclude that $Z^{(1)}(z, \bar{z}) =  (1+z)^{2 i r \epsilon} (1+\bar{z})^{2 i r \epsilon} Z^{(\text{col})}(z, \bar{z})$. The same is valid for $Z^{(2)}$ and $Z^{(\text{row})}$, so in the end we obtain
\begin{equation}
Z^{S^2}_{1,2}(z, \bar{z}) = (1+z)^{2 i r \epsilon} (1+\bar{z})^{2 i r \epsilon} Z^{S^2}_{2,1}(z,\bar{z})
\end{equation}
Taking into account the appropriate normalizations, this implies
\begin{equation}
Z_{1,2}^{\text{norm}}(z, \bar{z}) = Z_{2,1}^{\text{norm}}(z,\bar{z}) \ \ .
\end{equation}
As further example, we will briefly comment about the $\mathcal{M}_{3,1}$ and $\mathcal{M}_{4,1}$ cases. 
For $\mathcal{M}_{3,1}$ there are three contributions to the partition function:
 
\hspace{0.5 cm}

\begin{tabular}{ccl}
${\tiny\yng(1,1,1)}$ & from the poles & $\lambda_{1} = -i a$, $\lambda_{2} = -i a -i \epsilon_{1}$, $\lambda_{3} = -i a -2 i \epsilon_{1}$ \\ 
${\tiny\yng(1,2)}$ & from the poles & $\lambda_{1} = -i a$, $\lambda_{2} = -i a -i \epsilon_{1}$, $\lambda_{3} = -i a - i \epsilon_{1}- i \epsilon_{2}$ \\ 
${\tiny\yng(3)}$ & from the poles & $\lambda_{1} = -i a$, $\lambda_{2} = -i a -i \epsilon_{2}$, $\lambda_{3} = -i a -2 i \epsilon_{2}$
\end{tabular} 

\hspace{0.5 cm}

\noindent The study of the vortex contributions tells us that there is an equivariant mirror map, which has to be inverted; however, this is taken into account by the normalization factor. Then, the $r$ expansion gives
\begin{eqnarray}
Z_{3,1}^{\text{norm}} &=& \dfrac{1}{r^4 (\E \EE)^2} \Big[ -\dfrac{1}{6 r^2 \E\EE} -\dfrac{1}{4}\ln^{2} (z \bar{z}) + ir(\E + \EE) \Big( -\dfrac{1}{12} \ln^{3} (z \bar{z}) - \ln (z \bar{z})(\text{Li}_2(z) + \text{Li}_2(\bar{z})) \nonumber\\
&+& 2 (\text{Li}_3(z) + \text{Li}_3(\bar{z})) + 3 \zeta(3) \Big)  \Big] 
\label{3}
\end{eqnarray}

\noindent For $\mathcal{M}_{4,1}$ we have five contributions:

\hspace{0.5 cm}

\begin{tabular}{ccl}
${\tiny\yng(1,1,1,1)}$ & from & $\lambda_{1} = -i a$, $\lambda_{2} = -i a -i \epsilon_{1}$, $\lambda_{3} = -i a -2 i \epsilon_{1}$, $\lambda_{4} = -i a -3 i \epsilon_{1}$ \\ 
${\tiny\yng(1,1,2)}$ & from & $\lambda_{1} = -i a$, $\lambda_{2} = -i a -i \epsilon_{1}$, $\lambda_{3} = -i a - 2 i \epsilon_{1}$, $\lambda_{4} = -i a - i \epsilon_{2}$ \\ 
${\tiny\yng(2,2)}$ & from & $\lambda_{1} = -i a$, $\lambda_{2} = -i a -i \epsilon_{1}$, $\lambda_{3} = -i a - i \epsilon_{2}$, $\lambda_{4} = -i a - i\epsilon_{1} - i \epsilon_{2}$ \\
${\tiny\yng(1,3)}$ & from & $\lambda_{1} = -i a$, $\lambda_{2} = -i a -i \epsilon_{2}$, $\lambda_{3} = -i a - 2 i \epsilon_{2}$, $\lambda_{4} = -i a - i \epsilon_{1}$ \\
 ${\tiny\yng(4)}$ & from & $\lambda_{1} = -i a$, $\lambda_{2} = -i a -i \epsilon_{2}$, $\lambda_{3} = -i a -2 i \epsilon_{2}$, $\lambda_{4} = -i a -3 i \epsilon_{2}$
\end{tabular} 

\hspace{0.5 cm}

\noindent Again, we normalize and expand in $r$ to obtain
\begin{eqnarray}
Z_{4,1}^{\text{norm}} &=& -\dfrac{1}{r^6 (\E \EE)^3} \Big[ - \dfrac{1}{24 r^2 \E \EE} -\dfrac{1}{8}\ln^{2} (z \bar{z}) + ir(\E + \EE) \Big( -\dfrac{1}{24} \ln^{3} (z \bar{z}) - \ln (z \bar{z})(\frac{1}{2}\text{Li}_2(z) + \frac{1}{2}\text{Li}_2(\bar{z})) \nonumber\\
&+& 2 (\frac{1}{2}\text{Li}_3(z) + \frac{1}{2}\text{Li}_3(\bar{z})) + \frac{3}{2} \zeta(3) \Big)  \Big] 
\label{4}
\end{eqnarray}
As we will discuss in the Appendix 
the resulting Gromov-Witten potentials for these cases are in agreement with the quantum multiplication in the Hilbert scheme of points obtained in \cite{OP}.

\section{Donaldson-Thomas theory and stringy corrections to the Seiberg-Witten prepotential}
\label{sec:DT}

It is very interesting to analyse our system also from the D5-brane dynamics viwepoint. This
is a six-dimensional theory which should be related to higher rank equivariant Donaldson-Thomas theory
on $\mathbb{C}^2\times \mathbb{P}^1$. Indeed an interesting and promising aspect is that for $N>1$ the D1 contributions to 
the D5 gauge theory dynamics do not factor in abelian $N=1$ terms and thus keep an intrinsic non-abelian nature, contrary to what happens
for the D$(-1)$ contributions in the Coulomb phase \cite{richard}. 

To clarify this connection, let us notice
that a suitable framework to compactify the Donaldson-Thomas moduli space 
was introduced in
\cite{D} via ADHM moduli sheaves. In this context one can show that ${\cal I}_{k,1}={\cal I}_{DT}$. 
Moreover the ${\cal I}_{k,1}$-function reproduces the 1-legged Pandharipande-Thomas vertex
as in \cite{PT} for the case of the Hilbert scheme of points of ${\mathbb C}^2$, while 
the more general ADHM case should follow as the generalization to higher rank.
The case of the Hilbert scheme of points is simpler and follows by \cite{OP2}.

The partition function of the D1-branes computed in the previous sections provides non perturbative corrections to the D5-brane dynamics. The vortices
represent the contribution of D(-1)-branes located at the North and South pole of the blown-up two-sphere. 
It is then natural to resum the D(-1)-D1-brane contributions as
\beq
Z^{DT}_N=\sum_{k}q^{2kN}Z_{k,N}^{hol}=\sum_{k,\beta} N_{k,\beta}q^{2kN} z^\beta
\label{resum}
\eeq
where $q=e^{2\pi i \tau}$ is the D1-brane counting parameter and $z$ the D(-1)-brane one. 
In the second equality we considered the expansion $z$ of the holomorphic part of the 
spherical partition function, where $\beta\in H_2({\cal M}_{k,N},\mathbb{Z})$.

It is interesting to study the free-energy of the above defined partition function and its reductions in the 
four dimensional blow-down limit $r\to 0$.     
Indeed, let us observe that the D5 brane theory in this limit 
is described by an effective four-dimensional ${\cal N}=2$ supersymmetric gauge theory at 
energies below the UV cutoff provided by the inverse radius of the blown-up sphere $1/r$ \cite{lerda}.
Comparing the expansion (\ref{resum}) to the results of Sec.2.1, we obtain that the former reduces to the
standard Nekrasov instanton partition function upon the identification $q=\Lambda r$. 
Moreover,
keeping into account the limiting behaviour as $\epsilon_i\sim 0$ we have just discussed in the previous subsection,
namely that $Z_{k,N}^{S^2}$ has the same divergent behaviour as $Z_{k,N}^{Nek}$ due to the
equivariant regularization of the $\mathbb{R}^4$ volume $\frac{1}{\epsilon_1\epsilon_2}$,
one can present the resummed partition function (\ref{resum}) in the form
\beq
Z_N^{DT}
={\rm exp}\left\{
-\frac{1}{\epsilon_1 \epsilon_2}{\cal E}(a,\epsilon_i,\Lambda;r,z)
\right\}
\label{eff}
\eeq
where ${\cal E}$ is the total free energy of the system and is a regular function as 
$\epsilon_i\sim0$. The effective geometry arising in the semiclassical limit $\E,\EE\to 0$ of \eqref{eff}
would provide informations about the mirror variety encoding the enumerative invariants in \eqref{resum}.

In order to pursue this program it is crucial to complement our analysis by including the perturbative sector
of the $N$ D5-brane theory in the geometry $\mathbb{C}^2\times T^* \mathbb{P}^1 \times\mathbb {C}$
whose world-volume theory is described at low-energy by a ${\cal N}=1$ super Yang-Mills theory in six dimensions
on $\mathbb{C}^2\times \mathbb{P}^1$.
Its perturbative contribution can be computed by 
considering the dimensional reduction  
down to the two-sphere. This gives rise to a ${\cal N}=(4,4)$ supersymmetric gauge theory, containing three chiral multiplets in the adjoint representation
with lowest components $({Z}_i, \Phi)$, $i=1,2$, where $Z_1,Z_2$ and $\Phi$ describe the fluctuations along $\mathbb{C}^2$ and $\mathbb{C}$ respectively.
Around the flat connection, the vacua are described by covariantly constant fields $D_{\adj(\Phi)}Z_i=0$ satisfying
\beq
[Z_1,Z_2] = 0
\eeq
The Cartan torus of the rotation group acts as $(Z_1,Z_2)\to(e^{-\E}Z_1,e^{-\EE} Z_2)$
preserving the above constraints.
The one-loop fluctuation determinants for this theory
are given by
\beq
\frac{\det(D_{\adj (\Phi)})\det(D_{\adj (\Phi)}+ \E + \EE ) }
{\det(D_{\adj (\Phi)}+\E)\det(D_{\adj (\Phi)}+\EE) }
\ \ .
\eeq
The zeta function regularization of the above ratio of determinants
reads
\beq
\exp\left[- \frac{d}{ds} \frac{1}{\Gamma(s)}\int_0^\infty  \frac{dt}{t^{1-s}} \ \tr \ e^{t D_{\adj(\Phi)}}(1-e^{\E t})(1-e^{\EE t})\right]_{s=0}
\eeq
which can be seen as the regularization of the infinite product
\beq
\prod_{j,k}^\infty \prod_{l\neq m} \left(
\frac{\Gamma\left(1-ir(a_{lm}- j\E - k\EE)\right)}{ir\Gamma\left(ir(a_{lm} - j\E - k\EE)\right)}
\right)^{-1}
\label{gammif}
\eeq
The above formula is a deformation of the standard formula expressing the perturbative part of the 
Nekrasov partition function 
\beq
Z_{Nek}^{Pert}=\prod_{l\neq m}\prod_{j,k\geq 1} X_{lm,j,k}^{-1}=\prod_{l\neq m}\Gamma_2(a_{lm};\E,\EE)
\label{Npert}
\eeq
with $X_{lm,j,k}=a_{lm}-j\E-k\EE$, in terms of Barnes double $\Gamma$-function \cite{NO} (see also \cite{LMN}).
Eq.\eqref{gammif} is obtained by resumming 
the Kaluza-Klein modes on the two-sphere over each four dimensional gauge theory mode
organized in spherical harmonic $SU(2)$ multiplets.
This can be done by applying the methods in \cite{BC} to each tower before boson/fermion cancellation.
More details on the derivation can be found in Appendix \label{D5}.
Summarizing, the D5-D5 partition function is then given by
\beq
Z_{D5-D5}^{S^2}=
\prod_{l\neq m}
\Gamma_2(a_{lm};\E,\EE)\frac{\Gamma_3\left(a_{lm};\E,\EE,\frac{1}{ir}\right)}
{\Gamma_3\left(a_{lm};\E,\EE,-\frac{1}{ir}\right)}
\label{55}
\eeq
and implements the finite $r$ corrections to the perturbative Nekrasov partition function.
The equality in (\ref{55}) follows by regularizing the infinite set of 
poles of the ratio of $\Gamma$ functions. 
Indeed by using the standard properties of the $\Gamma$-function is it easy to see that \eqref{55}
reduces in the $r\to0$ limit to \eqref{Npert} plus corrections expressible in power series
in $r$ and $\E,\EE$.
More detailed calculations of the first terms of this expansion are presented in Appendix C.

We thus conclude that 
in the limit $r\to 0$, ${\cal E}\to {\cal F}_{Nek}$ the Nekrasov prepotential of the 
${\cal N}=2$ gauge theory in the $\Omega$-background. Therefore for $r\to 0$ the effective geometry
arising in the semiclassical limit of \eqref{eff} is the Seiberg-Witten curve of 
pure ${\cal N}=2$ super Yang-Mills \cite{NO}. 
Higher order corrections in $r$ to this geometry
encode the effect of stringy corrections. Indeed,
the total free energy contains additional world-sheet corrections in $z$ and therefore
$$
{\cal E}= {\cal F}^{Nek}(a,\epsilon_i,\Lambda) + {\cal F}^{stringy}(a,\epsilon_i,\Lambda;r,z)
$$
These are genuine string corrections to the ${\cal N}=2$ gauge theory in the $\Omega$-background
describing the finite radius effects of the blown-up sphere resolving the $A_1$ orbifold 
singularity.
Let us notice that ${\cal F}^{stringy}$ is higher order in the $\epsilon_i$ expansion
with respect to ${\cal F}^{Nek}$, therefore, in this scaling scheme, the resulting Seiberg-Witten limit
$\lim_{\epsilon_i\to 0}{\cal E}={\cal F}^{SW}$ is unchanged.

As we discussed in the previous section, the stringy contributions
are given by a classical term describing the equivariant classical intersection
theory in the ADHM moduli space and a world-sheet instanton contribution describing its 
quantum deformation, that is
\beq
{\cal F}^{stringy}(a,\epsilon_i,\Lambda;r,z)={\cal F}^{stringy}_{cl}(\epsilon_i;r,z)
+
\epsilon{\cal F}^{stringy}_{ws}(a,\epsilon_i,\Lambda;r,z).
\eeq

Following \cite{NS}
we can consider the effect of a partial $\Omega$-background by studying the 
limit $\epsilon_2\to 0$ in the complete free energy.
Defining 
\beq
{\cal V}={\rm lim}_{\epsilon_2\to 0}\frac{1}{\epsilon_2}{\rm ln} Z_N^{DT}
\eeq
we find that
\beq
{\cal V}={\cal W}^{NS} + {\cal W}^{stringy}
\eeq
where ${\cal W}^{NS}$ is the Nekrasov-Shatashvili twisted superpotential of the reduced two dimensional 
gauge theory and ${\cal W}^{stringy}$ are its stringy corrections.
According to \cite{NS}, ${\cal W}_{NS}$ can be interpreted as the Yang-Yang function 
of the quantum integrable Hitchin system on the M-theory curve (the sphere with two maximal 
punctures for the pure ${\cal N}=2$ gauge theory).
The superpotential ${\cal V}$ should be related 
to the quantum deformation of the relevant integrable system underlying the classical Seiberg-Witten geometry \cite{MO}.

\section{Conclusions}
\label{sec:conclusion}

In this paper we considered the dynamics of a D1-D5 system on a resolved $A_1$ singularity.
We calculated exactly the partition function of the GLSM describing the D1 dynamics and found
that this provides finite $S^2$-size corrections to the Nekrasov instanton partition function.
We showed that these corrections describe the quantum cohomology of the ADHM moduli space by identifying
the vortex partition functions appearing in the supersymmetric localization of the D1 partition function
with Givental's ${\cal I}$-function. A more detailed account on the identification between vortices
and equivariant Gromov-Witten invariants will be presented in \cite{BSTV}.
By using these results, we
proposed a contour integral formula for the Givental ${\cal I}$-function of the ADHM moduli space.
This provide a conjectural explicit formula for the quasi-map ${\cal I}$-function defined in
\cite{CKM}.
In the case of the Hilbert scheme of points our results match those of \cite{OP,Fontanine}. 
This suggests to push further the comparison between our approach 
and that of \cite{MO}
based on quantum deformed integrable systems.
In particular, the 
Yangian action on the quantum cohomology should be also realized on the ${\cal I}_{k,N}$-function.
This analysis would also be relevant in order to gain insights on a possible AGT counterpart \cite{AGT} 
of our results.
We observe that the finite size corrections vanish in the limit $\epsilon=\E+\EE \to 0$; this is 
consistent with the results of \cite{MO}. From the string-theoretic viewpoint, we notice that
this limit leads to an anti-self-dual $\Omega$-background and corresponds to a supersymmetry enhancement whose mark point is complete boson-fermion
cancellation of the one-loop determinants. More precisely, for $\epsilon=0$ the Nekrasov partition function we compute is interpreted in terms of type IIB 
superstring amplitudes in graviphoton background. These are known to decouple from the K\"ahler modulus of the resolved two-sphere which sits in
a hypermultiplet. The fact that for $\epsilon\ne 0$ we find non trivial corrections indicates that a world-sheet interpretation of the Nekrasov
partition function should be given in terms of (IIB) superstring amplitudes that couple also to hypermultiplet moduli. 
Our results should also follow from more traditional world-sheet techniques. The first finite $r$ corrections
to the $D=4$ gauge theory should be computed by the disk amplitudes with insertions of the 
string vertex corresponding to the blow-up mode.
At finite $r$ one should be able to treat the open string computation on the resolved geometry.

We discussed also the D5-brane viewpoint and its relation to higher rank equivariant Donaldson-Thomas (DT)
on $\mathbb{C}^2\times\mathbb{P}^1$. Vortex counting on the $D1$-branes amount to consider a full $D5-D1-D(-1)$ system
with $D(-1)$ located at the North and South pole of the two-sphere.
We observe that for the rank one case a direct relationship with
the quantum cohomology of the Hilbert scheme of points of $\mathbb{C}^2$ was pointed out in \cite{OP,OP2,PT}.
A proposal for the description of higher rank DT was formulated 
in \cite{D} in terms of ADHM moduli sheaves. 
This is strictly related to our approach and it would 
be interesting to further analyse the relation between the two.  
Indeed, the mathematical counterpart of our approach to the D1-D5 system corresponds
to study the representations of the associated ADHM quiver in the abelian category of coherent sheaves
over $\mathbb{P}^1$ corresponding to a particular case of \cite{D}.

Let us remark that our results point toward the existence of an effective geometry
encoding these enumerative invariants. Indeed,
we observed that the finite size corrections do not affect the qualitative 
asymptotic behaviour as $\E,\EE\to 0$ of the D1-D5 partition function;
this allowed us to define a generalization of the Seiberg-Witten prepotential including effective world-sheet
instantons. It would be interesting to further analyse the effective geometry arising from this deformed prepotential
and its modular properties.
This should be related to a suitable deformation of the quantum Hitchin integrable system associated to the
four-dimensional gauge theory \cite{BT,morozov,mt,BMT-h,gorsky,Tan,TV}. 

Our approach can be extended in further directions:
\begin{itemize}
\item
one can enlarge the D-brane construction to include matter sectors by considering
D5 branes multicovering the $\mathbb{P}^1$. These correspond to the regular branes of the orbifold construction.
\item
one can replace $A_1$ by a general ADE singularity. The D1-D5 system in the corresponding resolved space provide
a brane engineering of ADE quiver gauge theories. Our approach gives an alternative route to obtain the results of \cite{NP}
and the quantum deformation \cite{chico} and extend them by including finite-size corrections.
\item
one can also consider the resolution of the geometry $\mathbb{C}^2/\Gamma \times T^*\mathbb{P}^1\times \mathbb{C}$ with
D5-branes along $\mathbb{C}^2/\Gamma \times \mathbb{P}^1$ and D1s wrapping the exceptional divisors.  
The D1-branes in this setting engineer the moduli space of instantons
in supersymmetric gauge theory on ALE space \cite{dconf}. The D1 partition function
computes the quantum cohomology of Nakajima quiver varietes and the D1-D5 system probes DT invariant on $ADE\times\mathbb{P}^1$. This amounts to study
the corresponding moduli space of sheaves over $\mathbb{P}^1$ \cite{private}.   
This system would compute finite size corrections to a quiver ${\cal N}=2$ gauge theory on the ALE space \cite{ale}
in terms of local DT theory on $\mathbb{C}^2/\Gamma \times \mathbb{P}^1$. The case of the Hilbert scheme of points
was studied in \cite{maulik-oblomkov}.
\item
one can consider D5 branes wrapping more general (resolved) Calabi-Yau singularities.  
The D1-D5 system would compute higher rank DT invariants for these spaces. One interesting class is given
by the non-commutative resolutions and their moduli space of quivers \cite{balazs,richard}. 
\item
a more general system of intersecting D5-branes can be considered where some of the D5 branes fill the whole resolved ADE
singularity and a transverse complex line.
This introduces surface operators in the gauge theory \cite{braverman} and 
provides a set-up to compute the quantum cohomology
of their moduli spaces, such as for example Laumon spaces \cite{laumon} and partial flag varieties \cite{BFRF}. 
Our approach should be compared with the results of \cite{BMO}. 
Furthermore, we remark that the above constitutes a useful set-up
to study the AGT correspondence \cite{aggtv,KT,Tan}.
\item
one can promote our calculations at the K-theoretic level by considering an uplift to M-theory.
We expect in this case a direct link to the K-theoretic Givental functions \cite{K-give} as discussed 
in \cite{DGH,BDP} for the $\mathbb{P}^1$ target space case.
This has applications to the algebra of Wilson loop operators in Chern-Simons theory \cite{kapu},
as we will discuss in \cite{BSTV}, and provide moreover a direct link to K-theoretic DT theory.
An interesting observation \cite{sara,BC,BDP,taki} is that gauge theories on  
squashed $S^3_b$ or $S^2\times S^1$ can be computed via different gluings of K-theoretic vortex partition
functions. We expect that these could be interpreted in terms of topological membrane theory \cite{membrane}.
\item
another challenging direction concerns the higher genus extension of supersymmetric localization to describe
D-branes wrapping general Riemann surfaces. We expect this to provide a cohomological field theory approach to
compute the higher genus quasi-maps of the relevant quiver \cite{CKM}.
\item
we remark finally that although we focused in this paper on unitary groups, our approach can be applied
to other classical gauge groups.

\end{itemize}

\section*{Acknowledgements}
We thank
F. Benini, A. Brini, S. Cremonesi, D.~Diaconescu, J.~Gomis, C. Kozcaz, A. Lerda, D. Maulik, K. Narain, A. Okounkov, S. Pasquetti, F. Perroni
and R. Szabo for interesting discussions and comments. We also thank the referee for her/his suggestions which led us to improve the presentation 
of the results.
This research was partly supported by the INFN Research Project PI14 ``Nonperturbative dynamics of gauge theory", 
by the INFN Research Project TV12, 
by   PRIN    ``Geometria delle variet\`a algebriche"
and
by  MIUR-PRIN contract 2009-KHZKRX

\appendix

\section{Equivariant quantum cohomology of ${\cal M}_{k,1}$ in the oscillator formalism}

Following the notation of \cite{BG} and \cite{OP},
the Fock space description of the equivariant cohomology of the Hilbert scheme of points of $\mathbb{C}^2$
is given in terms of creation-annihilation operators $\alpha_k$, $k\in \mathbb{Z}$
obeying the Heisenberg algebra
\beq
[\alpha_p,\alpha_q] = p\delta_{p+q}
\label{o}
\eeq
The vacuum is annihilated by the positive modes 
\beq
\alpha_p |\emptyset\rangle = 0 \ \ , p>0
\eeq
and the natural basis of the Fock space is given by
\beq
|Y\rangle = \frac{1}{|Aut(Y)|\prod_i Y_i}\prod_i \alpha_{Y_i}|\emptyset\rangle
\label{basis}
\eeq
where $|Aut(Y)|$ is the order of the automorphism group of the partition and $Y_i$
are the lengths of the columns of the Young tableau $Y$.
The total number of boxes of the Young tableau is counted by  
the eigenvalue of the energy $K=\sum_{p>0}\alpha_{-p}\alpha_p$.
Fix now the subspace ${\rm Ker}(K-k)$ for $k\in\mathbb{Z}_+$
and allow linear combinations with coefficients being rational functions of the 
equivariant weights. This space is then identified with the equivariant cohomology
$H^*_T\left({\cal M}_{k,1},\mathbb{Q}\right)$.
More specifically 
\beq
|Y\rangle\in H^{2n-2\ell(Y)}_T\left({\cal M}_{k,1},\mathbb{Q}\right),
\eeq
where $\ell(Y)$ denotes the number of parts of the partition $Y$.

The generator of the small quantum cohomology is then given by the state
$|D\rangle=-|2,1^{k-2}\rangle$ which describes the divisor corresponding to 
the collision of two point-like instantons.

The operator generating the quantum product by $|D\rangle$ is given by
the quantum deformed Calogero-Sutherland Hamiltonian
\beq
H_D
\equiv
\left(\E+\EE\right)\sum_{p>0}\frac{p}{2}\frac{(-q)^p+1}{(-q)^p-1}\alpha_{-p}\alpha_p
+
\sum_{p,q >0}\left[\E\EE\alpha_{p+q}\alpha_{-p}\alpha_{-q}-\alpha_{-p-q}\alpha_{p}\alpha_{q}\right]
-
\frac{\E+\EE}{2}\frac{(-q)+1}{(-q)-1} K
\label{ham}
\eeq
We can then compute the basic three point function as $\langle D|H_D|D\rangle$, where the 
inner product is normalized to be
\beq
\langle Y|Y'\rangle = \frac{(-1)^{K-\ell(Y)}}{\left(\E\EE\right)^{\ell(Y)}
|Aut(Y)|\prod_i Y_i}\delta_{YY'}
\label{form}
\eeq
The computation gives
$$
\langle D|H_D|D\rangle=(\E+\EE)\left(\frac{(-q)^2+1}{(-q)^2-1}-\frac{1}{2}\frac{(-q)+1}{(-q)-1}\right)
\langle D|\alpha_{-2}\alpha_2|D\rangle
=
(-1)(\E+\EE)\frac{1+q}{1-q}\langle D|D\rangle ,
$$
where we have used $\langle D|\alpha_{-2}\alpha_2|D\rangle=2\langle D|D\rangle$. By (\ref{form}), we finally get
\beq
\langle D|H_D|D\rangle=\frac{\E+\EE}{\left(\E\EE\right)^{k-1}}\frac{1}{2(k-2)!}
\left(1+2\frac{q}{1-q}\right)
\eeq
Rewriting $1+2\frac{q}{1-q}=\left(q\partial_q\right)^3\left[\frac{\left({\rm ln}q\right)^3}{3!}+2{\rm Li}_3(q)\right]$, 
we obtain that the genus zero prepotential is
\beq
F^0=F^0_{cl}+\frac{\E+\EE}{\left(\E\EE\right)^{k-1}}\frac{1}{2(k-2)!}\left[\frac{\left({\rm ln}q\right)^3}{3!}+2{\rm Li}_3(q)\right]
\eeq
The above formula precisely agrees with the results of Sect.3, see \eqref{3} and \eqref{4} for the cases $k=3,4$ respectively.

\section{Multi-instantons in the higher rank case}

Let us consider $\mathcal{M}_{2,2}$. In this case, five Young tableaux are contributing:
 
\hspace{0.5 cm}

\begin{tabular}{ccl}
$({\tiny\yng(1,1)}\,,\,\bullet)$ & from the poles & $\lambda_{1} = -i a_1$, $\lambda_{2} = -i a_1 -i \epsilon_{1}$ \\ 
$({\tiny\yng(2)}\,,\,\bullet)$ & from the poles & $\lambda_{1} = -i a_1$, $\lambda_{2} = -i a_1 -i \epsilon_{2}$ \\ 
$(\bullet\,,\,{\tiny\yng(1,1)})$ & from the poles & $\lambda_{1} = -i a_2$, $\lambda_{2} = -i a_2 -i \epsilon_{1}$\\ 
$(\bullet\,,\,{\tiny\yng(2)})$ & from the poles & $\lambda_{1} = -i a_2$, $\lambda_{2} = -i a_2 -i \epsilon_{2}$ \\ 
$({\tiny\yng(1)}\,,\,{\tiny\yng(1)})$ & from the poles & $\lambda_{1} = -i a_1$, $\lambda_{2} = -i a_2$
\end{tabular} 

\hspace{0.5 cm}

\noindent The order $r$ coefficient in the expansion of the various vortex functions is zero, so there is no equivariant mirror map to be inverted. As normalization, we will choose the simplest one, that is we multiply by a factor
\begin{equation}
(z \bar{z})^{i r (\epsilon_{1} + \epsilon_{2} - a_1 - a_2)}\left(\dfrac{\Gamma(1-i r \epsilon_{1})\Gamma(1-i r \epsilon_{2})}{\Gamma(1+i r \epsilon_{1})\Gamma(1+i r \epsilon_{2})}\right)^4
\end{equation} 
The expansion then gives 
\begin{eqnarray}
 Z_{2,2}^{\text{norm}} &=& \dfrac{1}{r^6 (\epsilon_1 \epsilon_2)^2 ((\epsilon_1 + \epsilon_2)^2 - (a_{1}-a_{2})^2)} \Big[ \dfrac{8 (\epsilon_1 + \epsilon_2)^2 + \epsilon_1 \epsilon_2 -2(a_{1}-a_{2})^2 }{r^2 ((2\epsilon_1 + \epsilon_2)^2 - (a_{1}-a_{2})^2)((\epsilon_1 + 2\epsilon_2)^2 - (a_{1}-a_{2})^2)}\nonumber\\
&&+\dfrac{1}{2}\ln^{2} (z \bar{z}) -i r(\epsilon_1 + \epsilon_2) \Big( -\dfrac{1}{6} \ln^{3} (z \bar{z}) -\ln (z \bar{z})(2\text{Li}_2(z) + 2\text{Li}_2(\bar{z})) \nonumber\\
&&+ 2 (2\text{Li}_3(z) + 2\text{Li}_3(\bar{z})) + c(\epsilon_i , a_i) \zeta(3) \Big)  \Big] 
\end{eqnarray}
where
\begin{equation}
c(\epsilon_i , a_i) = 8- \dfrac{4 \epsilon_1 \epsilon_2(\epsilon_1 \epsilon_2 + 2 (\epsilon_1 + \epsilon_2)^2 + 4 (a_1 - a_2)^2 )}{((2\epsilon_1 + \epsilon_2)^2 - (a_{1}-a_{2})^2)((\epsilon_1 + 2\epsilon_2)^2 - (a_{1}-a_{2})^2)}
\end{equation}

\section{Perturbative sector of the D5-brane theory}
\label{D5}

The one-loop contribution of the D5-D5 partition function on $\Omega$-background can be calculated by making use of
the equivariant index theorem for the linearized kinetic operator of the quantum fluctuations in six dimensions.
The low-energy field theory on the D5-branes is given by (twisted) maximally supersymmetric Yang-Mills theory on $\mathbb{C}^2\times S^2$.
The relevant complex is the $\bar\partial$ Dolbeaux complex \cite{INV}
\beq
0\to \Omega^{(0,0)}\to\Omega^{(0,1)}\to\Omega^{(0,2)}\to 0
\label{complex}
\eeq   
The equivariant index of the above complex is given by
\beq
\frac{(1-t_1^{-1}-t_2^{-1}+t_1^{-1}t_2^{-1})}{(1-t_1^{-1})(1-t_2^{-1})(1-t_1)(1-t_2)}\left(-\frac{t_3}{(1-t_3)}\right)\sum_{l,m} e^{ia_{lm}}
\label{index}
\eeq
where we used K\"unneth decomposition of the cohomology groups of $\mathbb{C}^2\times S^2$. The first factor
computes the equivariant index of the $\bar\partial$ operator on $\mathbb{C}^2$, the second that of $S^2$, while the third factor
the twisting by the gauge bundle in the adjoint representation. From \eqref{index} one can easily compute the ratio of determinants
of the one-loop fluctuations via the substitution r\^ule relating the equivariant index with the equivariant Euler characteristic of the complex:
\beq
\sum_{\alpha} c_\alpha e^{w_\alpha} \quad \to \quad \prod_\alpha w_\alpha^{c_\alpha}
\eeq
where $w_\alpha$ are the weights of the equivariant action and $c_\alpha$ their multiplicities. 
Here $t_1=e^{i\E}$, $t_2=e^{i\EE}$ $t_3=e^{i\EEE}$ with $\EEE=\sqrt{-1}/r$.

In order to extract the above data from Eq.(\ref{index}), we expand the $\mathbb{C}^2$ factor as
\beq
\sum_{i,j,\bar i,\bar j =0}^\infty \left(1-t_1^{-1}-t_2^{-1}+t_1^{-1}t_2^{-1}\right)t_1^{i-\bar i}t_2^{j-\bar j}
\label{c1}\eeq
and the $S^2$ factor in the two patches as
\beq
-\sum_{k=0}^\infty t_3^{1+k}
\label{c21}\eeq
at the north pole and as
\beq
\sum_{k=0}^\infty \left(t_3^{-1}\right)^k
\label{c22}\eeq
at the south pole.
Then the product of the eigenvalues is given by
\bea
\prod_{i,j,\bar i,\bar j =0}^\infty &&
\frac{\Gamma(a_{lm}+\E(i-\bar i)+\EE(j-\bar j)+\EEE)}{\Gamma(1-a_{lm}-\E(i-\bar i)-\EE(j-\bar j)+\EEE)}
\left(\frac{\Gamma(a_{lm}+\E(i-\bar i-1)+\EE(j-\bar j)+\EEE)}{\Gamma(1-a_{lm}-\E(i-\bar i-1)-\EE(j-\bar j)+\EEE)}\right)^{-1}\\
&&
\left(\frac{\Gamma(a_{lm}+\E(i-\bar i)+\EE(j-\bar j-1)+\EEE)}{\Gamma(1-a_{lm}-\E(i-\bar i)-\EE(j-\bar j-1)+\EEE)}\right)^{-1}
\frac{\Gamma(a_{lm}+\E(i-\bar i-1)+\EE(j-\bar j-1)+\EEE)}{\Gamma(1-a_{lm}-\E(i-\bar i-1)-\EE(j-\bar j-1)+\EEE)}\nonumber\label{ciccia}
\eea
where we used the Weierstrass formula for the $\Gamma$ function performing the product over the index $k$ in (\ref{c21},\ref{c22}).
The above product simplifies then to eq.(\ref{gammif}) in the text.


The leading order term in the small $r$ expansion of (\ref{55}) is (\ref{Npert}).
The first non vanishing correction in the expansion can be computed by expanding
\beq
\ln\left[
\frac{\Gamma\left(1-ir X\right)}{\Gamma\left(1+ir X\right)}\right]= 
2\gamma i X r -\frac{2}{3}iX^3\zeta(3)r^3+O(r^5)
\label{expa}\eeq
in (\ref{gammif}), where $\gamma$ is the Euler-Mascheroni constant. 
Carrying the product to a sum at the exponent and using zeta-function regularization for the infinite sums,
one gets 
$$
\ln\left[\frac{Z_{D5-D5}^{S^2}}{Z_{Nek}^{Pert}}\right] = 
-\gamma i r
\frac{N(N-1)}{12}\epsilon
+\frac{1}{12}i\zeta(3)r^3
\left(
\sum_{l\neq n}a_{ln}^2 -\frac{N(N-1)}{30} (\E^2-\E\EE+\EE^2)
\right)\epsilon
+O(r^5)
$$
where the first term is a regularization scheme dependent constant. 
We see that the first correction affects the quadratic part of the prepotential implying a modification of 
the beta function of the theory which keeps into account the contributions
of the KK-momenta on the ${\mathbb P}^1$.



\end{document}